# The physics and metaphysics of the conceptuality interpretation of quantum mechanics*

*Diederik Aerts*
Center Leo Apostel for Interdisciplinary Studies, Vrije Universiteit Brussel, Brussels, Belgium
E-mails: diraerts@vub.be, diraerts@gmail.com

*Massimiliano Sassoli de Bianchi*
Center Leo Apostel for Interdisciplinary Studies, Vrije Universiteit Brussel, Brussels, Belgium
Laboratorio di Autoricerca di Base, 6917 Barbengo, Switzerland
E-mails: msassoli@vub.be, autoricerca@gmail.com

**Abstract.** Quantum mechanics has maintained over the years the reputation of being "the most obscure theory." It works perfectly well, but nobody seems to know why. It has been argued that the difficulty in understanding quantum theory is our failed attempt to force onto it a wrong conceptual scheme, wanting at all costs to think about the objects of the theory as, precisely, objects, i.e., entities having continuously actual spatiotemporal properties. This too restrictive spatiotemporal scheme is most probably at the heart of the problem, as also underlined by the Einsteinian revolution. So, what could be an alternative? Many thinkers have suggested that we must surrender to the fact that our physical world is one of immanent powers and potencies. Aristotle did so *ante quantum litteram*, followed by scholars like Heisenberg, Primas, Shimony, Piron, Kastner, Kauffman, de Ronde, just to name a few, including the authors, who were both students of Piron in Geneva. However, if on the one hand a potentiality ontology puts the accent on the processes of change, responsible for the incessant shifts between actual and potential properties, on the other hand it does not tell what these changes are all about. In other words, the metaphysical question remains of identifying the nature of the bearer of these potencies, or potentialities, and of the entities that can actualize them. It is the purpose of the present article to emphasize that the above question has found a possible answer in the recent *conceptuality interpretation of quantum mechanics*, which we believe offers the missing ontology and metaphysics that can make the theory fully intelligible, and even intuitive. In doing so, we will also emphasize the importance of carefully distinguishing the different conceptual layers that are contained in its explanatory edifice, as only in this way one can properly understand, and fully appreciate, the explanatory power it offers, without promoting undue reductionisms and/or anthropomorphizations.

# 1 Introduction

There is a famous apocryphal phrase attributed to Richard Feynman, saying that: "if you think you understand quantum mechanics, you don't understand quantum mechanics." Whether Feynman exactly used these words remains unclear, but he certainly used similar ones when he wrote (Feynman 1985): "I think I can safely say that nobody understands quantum mechanics." In the same vein, Werner Heisenberg recounts a conversation he had in Copenhagen with Wolfgang Pauli and Niels Bohr, in June 1952, where Bohr appears to have said that (Heisenberg 1971): "those who are not shocked when they first come across quantum theory cannot possibly have understood it."

The above quotes, and similar ones that were pronounced by eminent quantum physicists, should not be interpreted, however, in an impossibilist sense, i.e., in the sense that we should give up trying to understand quantum physics. On the other hand, we cannot fail to observe that this is what many physicists have precisely done. Also, quoting Sean Carroll (2019): "What's surprising is that physicists seem to be O.K. with not understanding the most important theory they have." So, one could argue that quantum mechanics has earned an undeserved reputation of being the only theory nobody understands. But think of relativity, do people truly understand why the speed of light is the same in every referential frame? Probably not (Aerts 2018, Aerts & Sassoli de Bianchi, 2023).

In principle, one could also say that if we go deep enough, then every physical theory presents at some point a challenge in terms of understanding its most fundamental aspects. A great example is Isaac Newton's famous *hypotheses non fingo* (I feign no hypotheses), in relation to the problem of gravity, when he wrote (Newton 1726): "I have not as yet been able to discover the reason for these properties of gravity from phenomena, and I do not feign hypotheses. For whatever is not deduced from the phenomena must be called a hypothesis; and hypotheses, whether metaphysical or physical, or based on occult qualities, or mechanical, have no place in experimental philosophy. In this philosophy, particular propositions are inferred from the phenomena and afterwards rendered general by induction."

The mystery of gravity that Newton wasn't willing, or able, to speculatively explain, was about how material bodies are able to attract each other at a distance, without a direct contact between them. Gravity presented to Newton the problem of a *spooky action at a distance*, impossible to explain through a sufficiently reasonable hypothesis, considering the worldview of his time. As we know, we had to wait his successor, Albert Einstein, to have the first truly powerful explanation (Einstein 1916). In a sense, Einstein solved the problem by eliminating the problem, i.e., by eliminating the gravitational force, as in *General Relativity* gravitation becomes an expression of the very geometry of spacetime, hence entities simply move along *geodetics*, which are the generalization of the notion of straight line to curved spacetime.

Twist of fate, Einstein eliminated the apparently unexplainable spooky action at a distance of his predecessor just to be confronted with another spooky action, associated with the phenomenon of *quantum entanglement*. In a letter to Max Born, he famously wrote about quantum mechanics in the following terms (Born 1947): "I cannot seriously believe in it because the theory cannot be reconciled with the idea that physics should represent a reality in time and space, free from spooky action at a distance." This passage is particularly important, as it points to the main difficulty quantum mechanics confronts each person who really tries to understand its content: our preconception that physical reality is something that should entirely happen in space, and of course also in time, and the fact that we, apparently, lack intuition about what a non-spatial reality would be.

This is certainly not the only explanatory gap that quantum mechanics confronts us with, but when we look carefully into the theory, from a foundational perspective, it becomes plausible that *non-spatiality* can be considered *the* explanatory gap on which all the other interpretational issues nestle, like *superposition*, *entanglement*, *complementarity*, *indiscernibility*, and even the *measurement problem*. That being said, when we speak of understanding a theory, one needs to distinguish two aspects. The first is what we might call its technical understanding. This requires becoming familiar with the experiments that made the theory necessary, and with its formal language, in particular its mathematical edifice, which in the case of quantum mechanics can take different forms, depending on the approach adopted, the standard one being based on a Hilbertian formulation, using unit vectors for the states, self-adjoint operators for the observables, and orthogonal projections for the properties.

Then, there is the other aspect: the interpretation of the theory. According to Tim Maudlin, an interpretation is what transforms a formalism into a full-fledged physical theory. Quoting from a recent interview (Maudlin 2019):

> There is no doubt that [...] there is a mathematical formalism that we know how to derive predictions from, and those predictions can be accurate to fourteen decimal places, but what a [...] physical theory is, is more than just a mathematical formalism with rules, it should specify a physical ontology, which means: tell me what exists in the physical world. Are there particles? Are there fields? Is there spacetime? And tell me about these things [...] and the problem is that [...] quantum theory isn't a theory in that sense, it is just a formalism, and then what people call "interpreting quantum theory" – which sounds like a funny thing to do cause you'd say, well, I have a theory, what is an interpretation? – what's called "interpreting quantum theory" is really the development of precise physical theories that make the same predictions, or nearly the same predictions, that you get out of this standard mathematical recipe [...].

Following Maudlin, what is usually called an interpretation is therefore the specification of both an ontology and a metaphysics. By ontology we mean a specification of the inventory of entities that exist out there (Broad 1923), and by metaphysics we mean a specification of what these entities are, i.e., what their possibly ultimate nature is (Quine 1948). A similar perspective is adopted by de Ronde (2014, 2017), in his representational realist program, emphasizing that a theory without an ontology and metaphysics (what he calls the conceptual component of the theory) is not even deemed to be called a physical theory, also adding that, ideally, they should be *read off* from the formalism itself.

We agree with Maudlin and de Ronde on the necessity of providing an interpretation for quantum mechanics, so that it can be considered a bona fide physical theory, and not just a set of performant prescriptions for making all sorts of predictions, and it is precisely the purpose of the present article to argue in favor of a specific ontology and metaphysics for the theory, hence an interpretation, allowing quantum mechanics to become fully understandable. Of course, we don't know if our interpretation is necessarily the correct one. Further data needs to be collected in its favor, but in our opinion, at the current stage of development, it represents a very serious candidate, and in a sense the only candidate truly capable of explaining what is still considered inexplicable today. We also believe it is so far the only candidate that can explain *all* quantum conundrums, and not just some of them. Last but not least, our interpretation meets the desiderata of being the result of a bottom-up ontology-metaphysics, resulting from an attentive observation of the behavior of the different physical entities, followed by the question: "What nature would be able of producing such behavior?"

The article is organized as follows. In Section 2, we describe the key phenomena that any interpretation of quantum physics must be able to satisfactorily explain. In Section 3, we recall the genesis of the conceptuality interpretation and its basic assumptions. In Section 4, we show how the latter can provide a compelling explanation for all the deepest quantum mysteries. Finally, in Section 5, we offer some concluding remarks.

## 2 Quantum features

In this section, we describe some of the key quantum features and emphasize the mysteries they represent:

(1) quantum superposition
(2) quantum measurement
(3) quantum entanglement
(4) quantum complementarity
(5) quantum indistinguishability

### 2.1 Quantum superposition

In standard quantum mechanics, *quantum superposition* is intimately incorporated in the formalism. Indeed, being the state space a Hilbert space, i.e., a vector space, every state, as a vector, can always be written as a linear combination of two or more states. Also, the Schrödinger equation being linear, these combinations remain compatible with the evolution laws: if two vector-states obey the Schrödinger equation, the same will be true for their linear combination, i.e., for their superposition. Note that the historical term "superposition" comes from the fact that when vector-states are interpreted as waves, their linear combination can be interpreted as a superposition of waves. A more precise name, however, would be *quantum combination*.

It is unclear if the superposition principle should be valid in all experimental contexts, i.e., if it is true that, given two vector-states, all possible complex linear combinations of them are always bona fide states, in which the entity in question can find itself in. One reason to doubt it, is the so-called measurement problem. If the evolution laws are linear, and the measuring apparatus is also treated in a full quantum mechanical way, then starting from an initial superposition state, by linearity this will produce a final state that is also a superposition state of the composite system formed by the measured entity plus the apparatus. But this is not what is typically observed in a laboratory, hence there appear to be situations where the superposition principle does not apply.

Quantum superposition is also partially inhibited when *superselection rules* are in force (Streater & Wightman 1964), i.e., when the Hilbert state space can be decomposed as a direct sum of orthogonal subspaces, meaning that certain superpositions of vector-states cannot be prepared. Also, quantum superposition does not apply to all states if the state space is augmented to include *operator-states as genuine states, in addition to the vector-states, also called density matrices*, or *density operators* (Aerts & Sassoli de Bianchi 2014). Indeed, a convex linear combination of operator-states is again an operator-state. However, if the operator-states in question are one-dimensional orthogonal projection operators, i.e., they are the operatorial version of vector-states, then the projection operator associated with a normalized linear combination of the latter will in general be different from the operator-state obtained by considering a convex linear combination of the former, as the so-called interference terms will be absent (Aerts & Sassoli de Bianchi 2016).

Having said that, despite possible limitations, quantum superposition is certainly a widespread phenomenon that has been observed in countless experimental situations, also in relation to relatively large physical entities, like organic molecules (Gerlich et al. 2011, 2013) and even micro-mechanical resonators (O'Connell et al. 2010). The way it is typically highlighted in a laboratory is through the *interference effects* it can produce. The paradigmatic example is that of the double-slit experiment (Feynman et al. 1964), where the experimental context is such that the state of the entity is describable as a linear combination

$$\psi = a_1\psi_1 + a_2\psi_2$$

where $\psi_1$ is the state corresponding to the situation where only slit-1 is open, $\psi_2$ is the state corresponding to the situation where only slit-2 is open,[†] and $|a_1|^2 + |a_2|^2 = 1$.

When such a superposition state is measured, by allowing the quantum entity (an electron, a photon, etc.) to interact with a screen detector, operating as a position-measurement apparatus, an interference pattern will be observed, when numerous observations are collected, revealing that the entity in question cannot be described as being either in state $\psi_1$, with probability $|a_1|^2$, or in state $\psi_2$, with probability $|a_2|^2$ (see Figure 1). Physicists sometimes describe the situation by saying that the quantum entity passes through both slits at the same time, like a spatial wave, but this is a wrong way to describe what happens. The correct statement is that the entity *potentially* passes through both slits at the same time. Indeed, at no time, before the entity is ultimately revealed by the detection screen, there is an actual presence in space of the quantum entity.

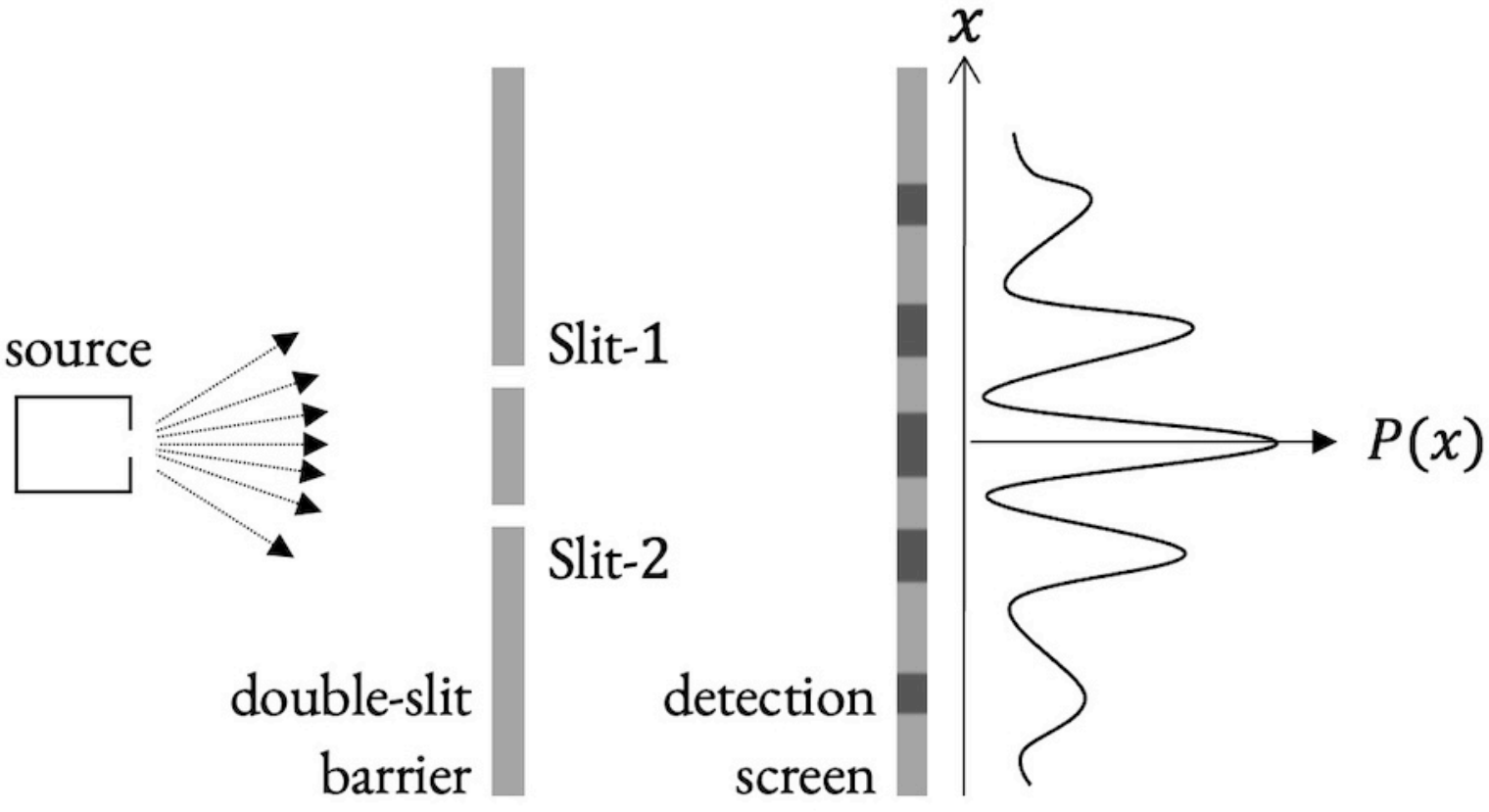

**Figure 1** In the double-slit experiment, when both slits are open, the probability distribution of impacts on the detection screen is not what one would expect if the entities emitted by the source were corpuscles, in that it corresponds to a fringe interference pattern, with the main fringe being at the center of the detection screen.

So, what is the real mystery of quantum superposition? Certainly not the fact that it describes a *potential mode of being*, as the notion of potentiality was already introduced by Aristotle to conceptualize the different possible processes of change, well before the advent of quantum mechanics (Aristotle 1995). The real mystery lies in the fact that these potential modes of being can be expressed in relation to mutually exclusive *spatial* properties. We know that an entity like an electron is not a wave, considering it can manifest as a localized spot on a detection screen. However, if it were only a spatially localized corpuscle, how could it potentially be present in both slits at the same time, without actually being present anywhere in space? In other words, what does it mean to be in a *non-spatial state*, i.e., in a state of *unactualized spatial properties*?

[†] Note that it is possible to decompose the wave function $\psi$ as a superposition of wave functions uniquely associated with single-slit situations only in regimes where the wavelengths become very small compared to the distances between the slits. See (Sassoli de Bianchi & Sassoli de Bianchi, 2025) and the references cited therein.

## 2.2 Quantum measurement

In addition to reversible and deterministic evolutions, for instance governed by the Schrödinger equation in the non-relativistic case, irreversible and indeterministic evolutions can also take place, when experimental procedures are carried out to observe specific physical quantities, called *quantum measurements*. These are processes of a *weighted symmetry breaking* kind, bringing a pre-measurement superposition state of the form

$$\psi = a_1\psi_1 + \cdots + a_N\psi_N$$

to only one of the $N$ possible outcome eigenstates $\psi_1, \cdots, \psi_N$, characterizing the measurement in question, assuming here for simplicity that their number $N$ is finite (see Figure 2). The process is usually called a *state reduction*, or *state collapse*, and although in a single measurement run it is impossible to predict its outcome, the transition probabilities, $p_1, \dots, p_N$, can still be calculated using the *Born rule*

$$p_i = |\langle\psi_i|\psi\rangle|^2 \qquad i = 1, \dots, N$$

expressing the statistical content of the theory and bringing it in contact with the experiments. More precisely, the transition $\psi \to \psi_1$ will be observed $p_1$% of the times, with $p_1 = |\langle\psi_1|\psi\rangle|^2$, the transition $\psi \to \psi_2$ will be observed $p_2$% of the times, with $p_2 = |\langle\psi_2|\psi\rangle|^2$, and so on, with $p_1 + \cdots + p_N = 1$.

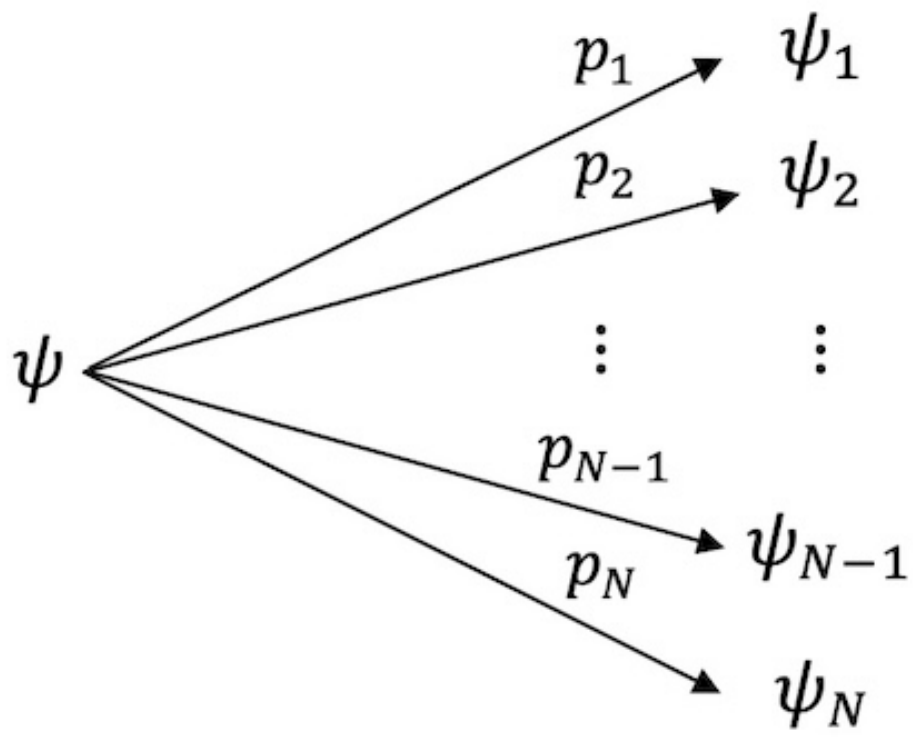

**Figure 2** In a quantum measurement, starting from a pre-measurement state $\psi$, several outcome-states, $\psi_1, \cdots, \psi_N$, are available to be actualized. Via the *Born rule*, the probabilities for the different possible outcomes, $p_1, \cdots, p_N$, can be calculated. These probabilities appear to be genuinely irreducible, i.e., to correspond to the maximum knowledge available to determine which outcome will ultimately be realized.

So, what is the mystery of a quantum measurement? Certainly not the fact that it describes a symmetry breaking process, as the basic idea of symmetry breaking is well-known, and uncontroversial, in different fields of physics. Think of the sudden and dramatic changes described in *bifurcation theory*, as part of the study of dynamical systems (Nicolis & Prigogine 1977). The notion of *bifurcation*, firstly introduced by Henri Poincaré (1885), precisely describes the paradigmatic situation where an equilibrium becomes increasingly unstable.

As an illustration, consider a pencil placed vertically on its tip, on a table. This is possible for as long as the contact area $A$ between the tip and the table (assuming here for simplicity that it is a flat circular surface) is non-zero, so that the gravity center of the pencil can orthogonally project inside that area, when the pencil is vertically positioned, in static equilibrium. The parameter $A$ plays here the role of the *order parameter*: when $A$ decreases, the environmental fluctuations will break the rotational symmetry (around its vertical axis) of the pencil's state, which by falling will acquire a non-rotationally symmetric state, actualizing a specific orientation, impossible to predict in advance (see Figure 3). Such orientation was only potential prior to the pencil's collapse; hence, the pre-collapse state can be described as a superposition state where the different directions are just potential elements of reality.

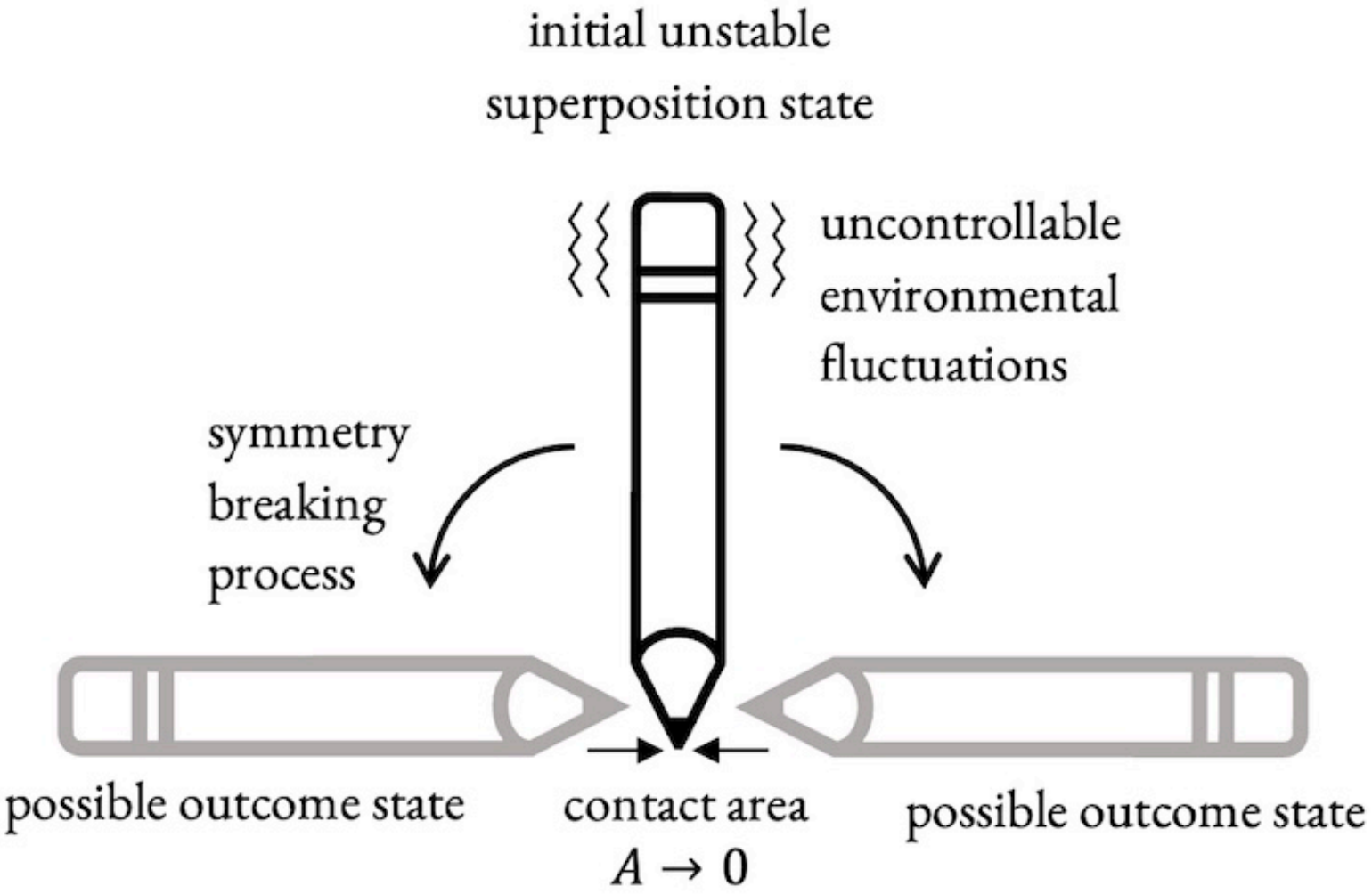

**Figure 3** A pencil placed vertically on its tip, on a table, is like a quantum superposition pre-measurement state, ready to collapse into one of many possible outcome-states, as the *contact area* order parameter tends to zero: $A \to 0$. In the initial, pre-measurement state, the different orientations are only potential, whereas a specific orientation is each time unpredictably actualized, when the pencil falls on the table.

So, why a quantum measurement is not viewed as a process like the pencil collapsing on the table? In our view, the main reason is that most physicists are still attached to the prejudice that a quantum entity, like any other physical entity, should be a spatial entity, so much so that, in the mind of many, *non-spatiality rhymes with non-reality*. In the above example of the pencil, adherence to this prejudice would be equivalent to consider that the only possible genuine states for the pencil are those where it lies flat on the table, i.e., those having a well-defined orientation on the tabletop plane, the table being in our example a metaphor for the Euclidean space. From our three-dimensional perspective, we obviously have no problem understanding that other states are possible for the pencil, beyond the tabletop. However, for quantum entities there is no "other place" available in which to reside, when they are in a state that corresponds to the superposition of different "place-states". So, as long as we associate *state* with *place*, we cannot understand what is going on, because a place is by definition contained somewhere within space. In Aerts & Sassoli de Bianchi (2015a), a similar metaphor was used in relation to the rolling of dice, where the table represented the place of residence of the classical spatial entities, i.e., the theatre in which, before the discovery of quantum physics, all physical entities were believed to belong.

In other words, if the spatial prejudice is maintained, it is not understandable why the outcome of a quantum position-measurement would be genuinely unpredictable, since it should be about the observation of a pre-existing spatial property. Therefore, when trying to unveil the mystery of quantum measurements, the first obstacle is that one usually does not accept that these processes could be viewed as (weighted) symmetry breaking processes, because this would in turn require to view the measured entities as non-spatial entities, and nobody knows what they would correspond to.

The second element of mystery in quantum measurements, which needs to be explained even if one accepts the idea that they are symmetry-breaking *contexts* that can bring non-spatial states into spatial states, is to understand what the nature of the uncontrollable fluctuations is that would cause the actualization processes, and how one can relate them to the predictions of Born's rule. In other words, the second element of mystery is about explaining what happens behind the scenes of a *quantum jump*.

## 2.3 Quantum entanglement

From a theoretical standpoint, *quantum entanglement* is a direct consequence of quantum superposition, when the latter is applied to composite systems. Indeed, if an entity $S$ is formed by two sub-entities, $S_A$ and $S_B$, with $S_A$ in the vector-state $\psi$, and $S_B$ in the vector-state $\varphi$, then the bipartite entity $S$ will be in the (tensor) *product state* $\psi \otimes \varphi$, which is again a vector-state. The situation where $S_A$ is in the vector-state $\varphi$ and $S_B$ is in the vector-state $\psi$ is of course also a possibility, corresponding to the product state $\varphi \otimes \psi$ for $S$. But if the superposition principle applies, the linear combination

$$a_1\psi \otimes \varphi + a_2\varphi \otimes \psi$$

is also a description of a possible state for $S$ (once duly normalized), called an *entangled state*, following a terminology introduced by Erwin Schrödinger.

Entangled states being a special case of superposition states, they present the same interpretative challenge the latter pose: that of understanding what non-spatiality truly is. They do so, however, in a more spectacular way, for the following reason. One can create experimental situations where the two entangled sub-entities, $S_A$ and $S_B$, are jointly measured in laboratories that are arbitrarily far away in space from one another. Following the prejudice saying that jointly executed measurements that are sufficiently spatially separated (hence, that are *spacelike* separated in a relativistic sense) should also be *experimentally separated*, one expects to only observe those correlations in the obtained statistics of outcomes that were already existing before the very execution of the joint measurements, i.e., correlations having their common causes in the past.

It was the merit of John Bell (1964) to derive mathematical inequalities involving experimentally accessible quantities that can be violated by quantum entangled entities, but not if the observed correlations can be explained as resulting from common causes in the past. In other words, Bell's inequalities allow for a demarcation between the correlations predicted by the quantum formalism, and the ordinary correlations of the "Bertlmann's socks kind" (Bell 1981, Aerts & Sassoli de Bianchi 2019). And many experiments performed in the eighties, then perfectioned over the years, confirmed the existence of quantum correlations, violating Bell's inequalities; see Bertlmann (1990) for a review.

So, what is the mystery of quantum entanglement? Certainly not the fact that it describes a situation where the observed correlations cannot be associated with common causes in the past. Indeed, also macroscopic composite systems can easily violate Bell's inequalities, when the sub-entities are *connected* in some way. This was emphasized by one of us already in the eighties of the last century, using a system formed by two *vessels of water* connected through a tube (see Figure 4), with the joint measurements so defined that correlations could be created in a contextual way, at each run of the joint measurements (Aerts 1982a, Aerts et al. 2019, Aerts & Sassoli de Bianchi 2019).

The question then arises: "Why quantum entanglement is not just viewed as a *connective element of reality*, like the tube connecting the two vessels of water?" The spatial prejudice is probably again the main obstacle here. Because then comes the additional question: "What would be the nature of a connective element of reality able to transform two spatially separated entities in a deeply interconnected whole and remain perfectly undetectable in the space between the two entities?" And there is also the further explanatory gap of understanding why quantum entanglement would be the default state: "Why quantum entities spontaneously entangle when they are allowed to interact, so that entanglement is ubiquitous in our physical reality, whereas it is also a relatively fragile state?"

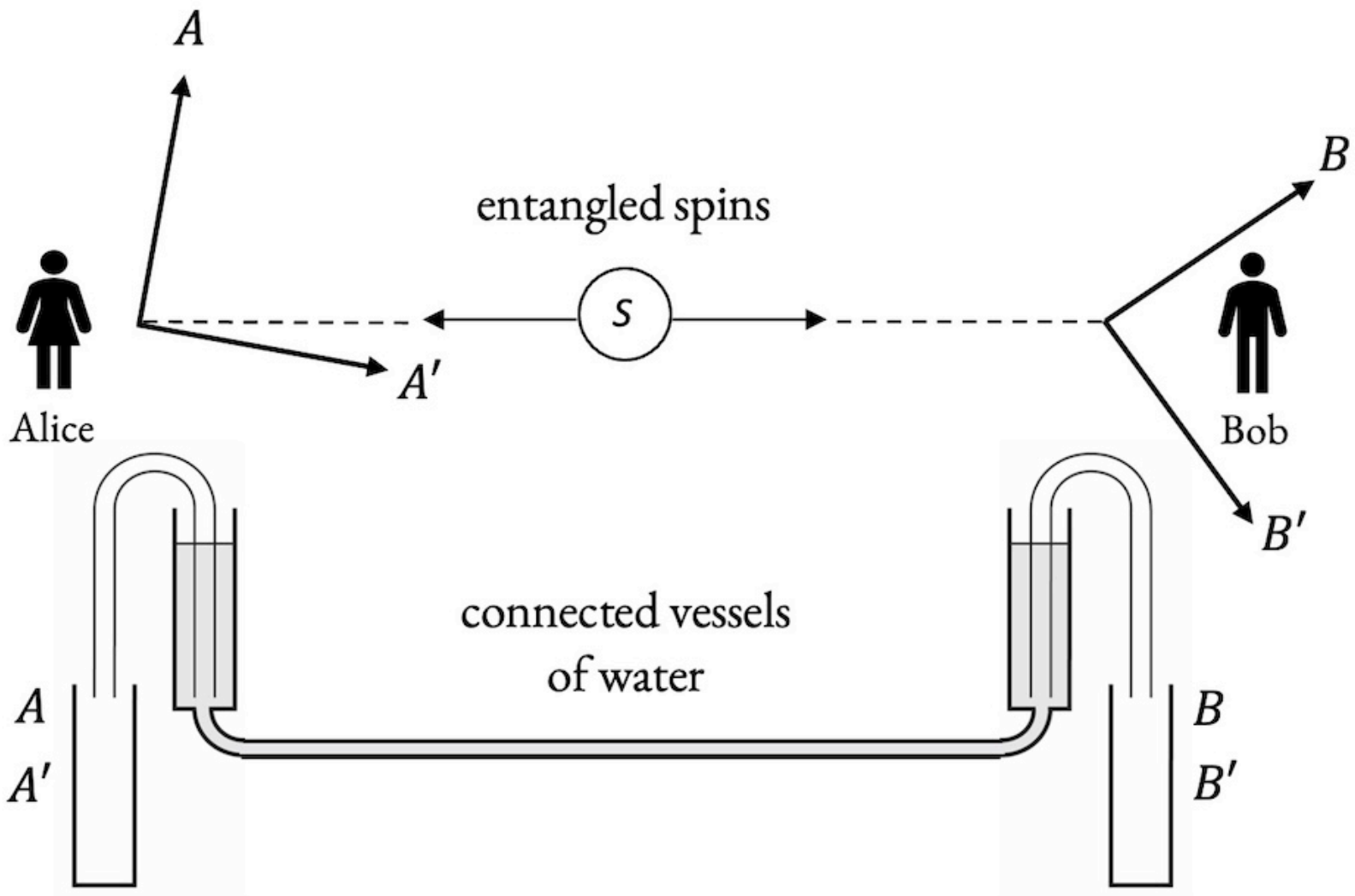

**Figure 4** Microscopic systems, like two entangled spins, and macroscopic systems, like two connected vessels of water, can both violate Bell-CHSH inequalities, when Alice's and Bob's joint experiments contextually create the observed correlations. For the two entangled spins, Alice's measurements, $A$ and $A'$, and Bob's measurements, $B$ and $B'$, are Stern-Gerlach measurements along different spatial directions. For the two connected vessels of water, they correspond to jointly extracting water from the vessels, using siphons, or checking its transparency (Aerts 1982a).

## 2.4 Quantum uncertainty

*Complementarity* is how Niels Bohr (1928) expressed Heisenberg's *uncertainty principle* in a broader conceptual framework. On one hand, one can say that complementarity expresses the fact that since quantum measurements are intrinsically invasive procedures, changing the state of the measured entity (except for when they are already in an eigenstate), some of them will be incompatible, in the sense that they cannot in general be carried out at the same time. If measurements are viewed as *interrogative processes*, this means that there are experimental questions that cannot be asked conjunctly, hence, that there are answers that cannot be received at the same time, the typical example being that of questions about *position* and *momentum* of a quantum entity.

So, what is the mystery of quantum uncertainty, or quantum complementarity? Certainly not the fact that it is impossible, in some circumstances, to carry out two measurements at the same time. Indeed, this does not imply that there wouldn't be a general way of testing if a quantum entity, say an electron, has the *meet property* of having a given position *and* a given momentum, at the same time. Because a property is just a *state of prediction* and a test is just a procedure that allows one to check the prediction in question, and tests associated with meet properties, called *product tests*, can always be defined (Piron 1976).

To use an example, which was introduced a long time ago in the doctoral dissertation of one of us (Aerts 1982b), consider a wooden cube and its two properties of *floating on water* and *burning well* (see Figure 5).

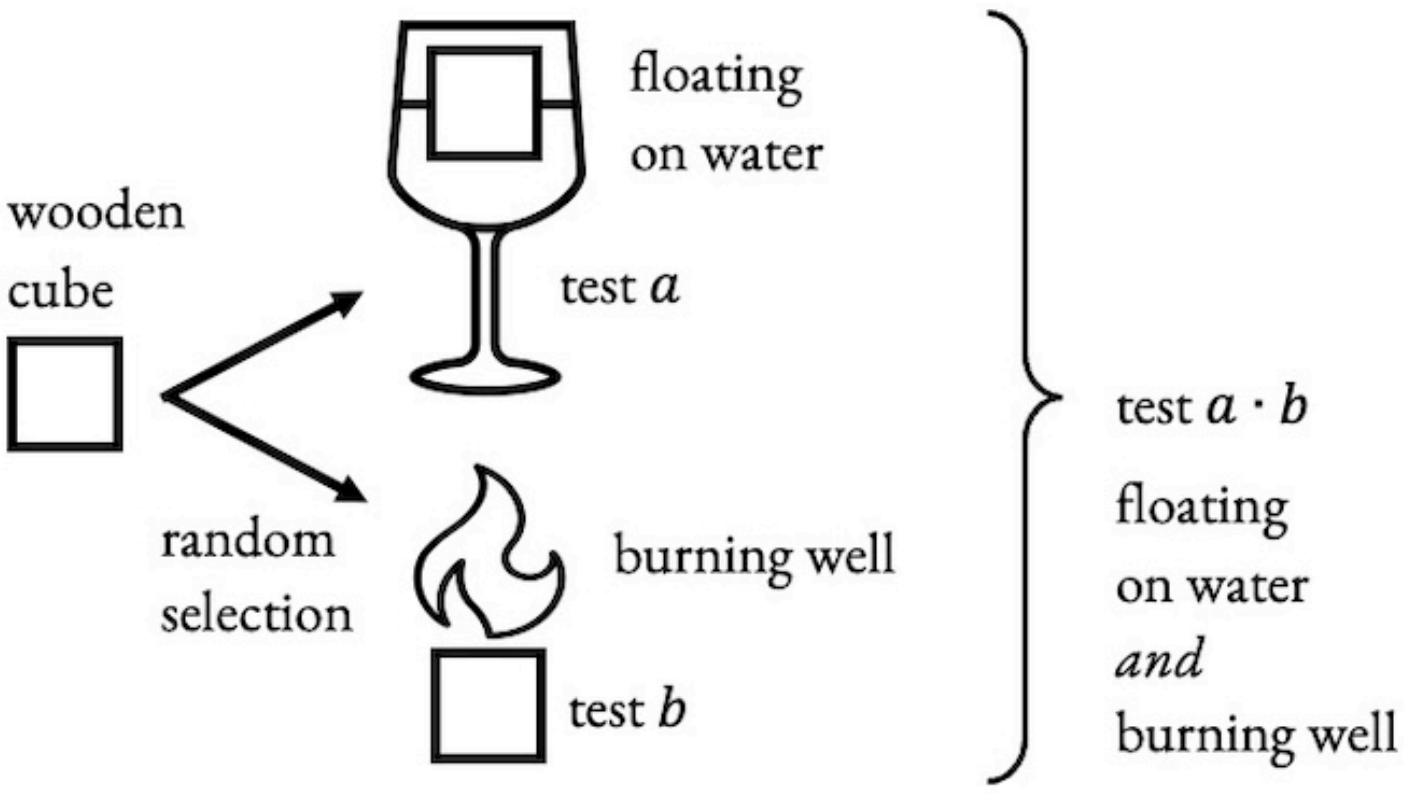

**Figure 5** The product test $a \cdot b$, for the meet property $A \wedge B$ of *floating on water and burning well*, consisting of randomly choosing, then executing, either test $a$ for property $A$ (floating on water) *or* test $b$ for property $B$ (burning well), then attributing the obtained result to $a \cdot b$.

Without going into details, the reader can easily understand why the burnability test is incompatible with the floatability test. In a nutshell, ashes do not float, and wet wood does not burn well. But one does not need to jointly perform the two tests to test the meet property of *floating on water and burning well*. Indeed, it is sufficient to randomly select one of the two tests and execute it, then consider the outcome (positive or negative) as the result of the product test itself. This is sufficient because the only way one can predict with certainty the positive outcome is to have both properties actual (the choice of which test to execute being by definition unpredictable). And we are all able to predict with certainty the positive outcome of such product test, compatibly with our prior knowledge that wooden cubes float on water and, also, burn well.

So, the experimental incompatibility of two properties of a given entity does not imply that they cannot be simultaneously actual, as the *disjunctive* logic of a product test indicates, and as the example of the wooden cube exemplifies. Why is this important? Because quantum uncertainty goes deeper than that, and this is where its mystery truly lies. Although one can easily define and carry out a product test for, say, an electron, aimed at assessing if its position and momentum observables have values within two given intervals, $\Delta x$ and $\Delta p_x$, along the $x$-direction, it is not possible, not even in principle, to predict with certainty the outcome of such product test, unless $\Delta x = \Delta p_x = \infty$.

In the quantum formalism, this is expressed by the fact that the position and momentum observables do not commute and that, because of that, states that would be jointly eigenstates of the two observables do not exist. Another way of saying this is that position and momentum measurements cannot be understood as sub-measurements of a bigger measurement, also associable with a self-adjoint operator. Therefore, the mystery behind quantum complementarity is again that of non-spatiality: the fact that classical spatial properties, like having a given position and a given momentum, cannot be jointly actualized, which is another way of saying that an elementary quantum entity is something very different from a corpuscle that would move along a specific trajectory in space.

Going more into the specifics of Heisenberg's uncertainty relations, there is an additional explanatory gap: that of being able to understand why the more a vector-state is peaked at about a given position in space, the less peaked that same vector-state is in momentum-space, in accordance with the properties of the *Fourier transform*. In other words, the more an entity is localized in position-space, the less it is localized in momentum-space, and Heisenberg's uncertainty relations famously express this *tradeoff* in quantitative terms, by means of a bound, for instance (Kennard 1927):

$$\Delta x \Delta p_x \geq \frac{\hbar}{2}$$

Note that a valid interpretation of the quantum formalism should be able to explain not only the above behavior, but also the existence of *reverse uncertainty relations* (Mondal et al. 2017), saying that in the same way that there are lower limits for the products (and sums) of variances of incompatible observables, also upper limits apply (albeit these are state-dependent and thus less fundamental) and this also needs to be explained. In other words, not only one cannot have states that are jointly maximally *sharp* in position and momentum, but in most situations of physical interest neither can we have states that are jointly maximally *unsharp* in position and momentum.

## 2.5 Quantum indistinguishability

*Quantum indistinguishability* is the impossibility of differentiating identical quantum entities. This means that in a multipartite system formed by $N$ identical and indistinguishable entities, no observable can allow one to detect if two entities have been permuted in the system, not even in principle. Consequently, only *symmetric Hermitian operators*, commuting with all possible permutations, can describe the observables.

There are two typologies of fundamentally indistinguishable entities in quantum mechanics:

(1) *fermions*, with a fractionary spin, obeying Pauli's exclusion principle, whose states are antisymmetric under permutations;
(2) *bosons*, with an integer spin, not obeying Pauli's exclusion principle, whose states are symmetric under permutations.

When they form assemblies, fermions are not counted in the same way as bosons, precisely because the former obey *Pauli's exclusion principle*, whereas the latter do not. More precisely, fermions obey the *Fermi-Dirac statistics* (Fermi 1926, Dirac 1926), whereas bosons obey the *Bose-Einstein statistics* (Bose 1924, Einstein 1924). Classical entities, on the other hand, being in principle distinguishable, their assemblies are not permutation invariant and obey the historical *Maxwell-Boltzmann statistics* (Maxwell 1860, Boltzmann 1877); see Figure 6.

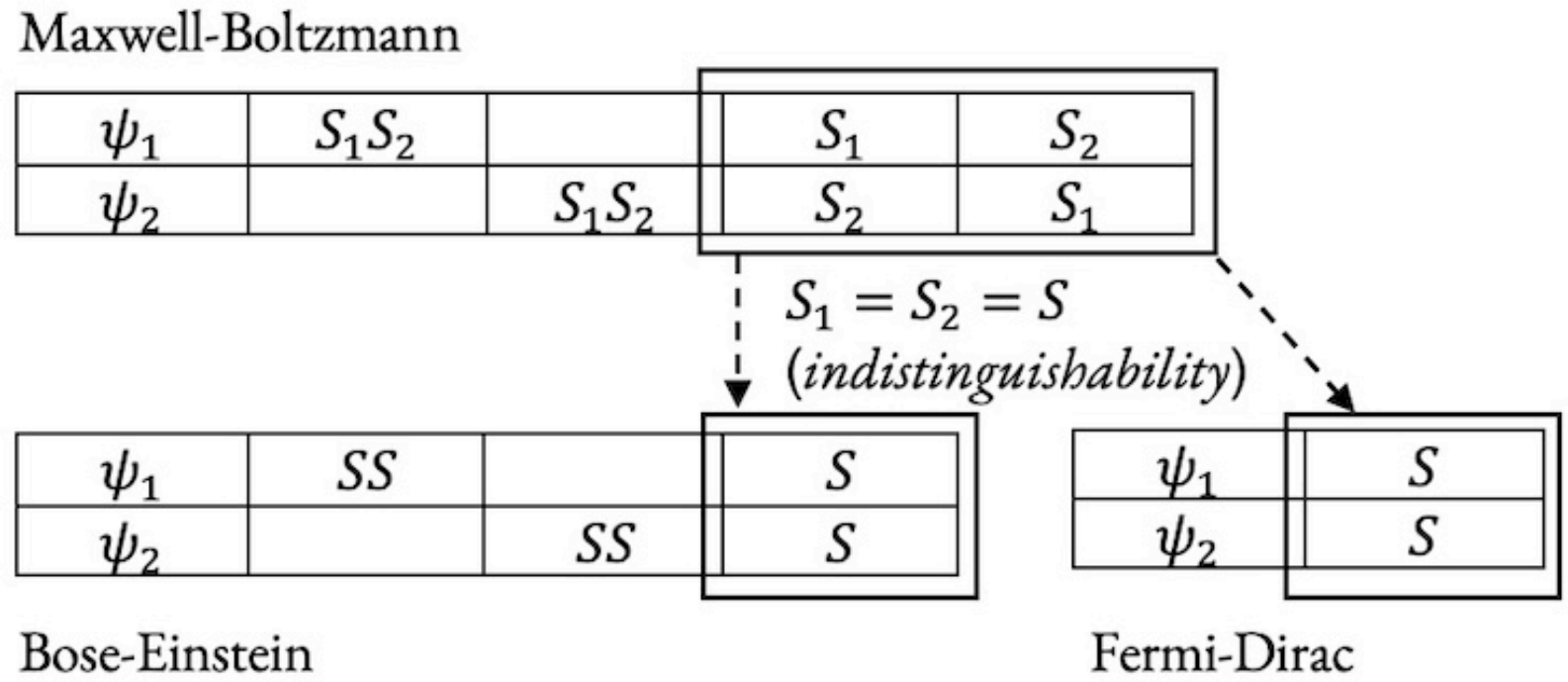

**Figure 6** If two entities, $S_1$ and $S_2$, can only be in two different states, $\psi_1$ and $\psi_2$, then, when considered as a system, the states it can be in depend on whether the two entities are distinguishable (Maxwell-Boltzmann way of counting) or indistinguishable. In the latter case, and additional distinction is whether $S_1$ and $S_2$ can be in the same state (Bose-Einstein way of counting) or not (Fermi-Dirac way of counting).

So, what is the mystery of quantum indistinguishability? Certainly not the fact that identical physical entities would not exist. We must not confuse the notion of *identicality* with that of *indistinguishability*. The latter requires the former but not the other way around. Two entities are said to be identical if they have the same intrinsic properties. For example, two electrons are identical because they have the same electric charge, rest mass, one-half fractionary spin, etc. But being identical does not mean being indistinguishable. Before the advent of quantum mechanics, two classical corpuscles, like two electrons, were considered identical but nevertheless distinguishable, for instance because one could use their trajectories to set them apart. Quantum entities, however, being non-spatial, one cannot use trajectories to distinguish them. Would this mean that the mystery of quantum indistinguishability is the same as that of quantum non-spatiality? Not exactly.

Indistinguishability due to lack of trajectory does not imply that two identical quantum entities cannot be distinguished: it simply means that one does not have a notion of trajectory to do so. This does not exclude, however, a possible distinguishability of identical entities by other means. In other words, indistinguishability due to lack of trajectory is a necessary but not sufficient condition to deduce quantum indistinguishability, which is a stronger statement. To put it differently, quantum indistinguishability is not equivalent to quantum non-spatiality, although quantum indistinguishability implies quantum non-spatiality, as spatiality implies distinguishability.

What is then the specific mystery of quantum indistinguishability? It is the fact that two identical quantum entities, when truly indistinguishable, they nevertheless can remain individuals. This appears to contradict *Leibniz's ontological principle* of the identity of the indiscernibles, stating that no two distinct entities that are exactly equivalent in all their properties (intrinsic and accidental). One might object that Leibniz's principle is at least not violated by fermions, because of Pauli's exclusion principle, but even in this case, if one wants to attribute individual states to, say, two electrons in an antisymmetric (entangled) state, the only way to do so is by generalizing the notion of state to also include density operators. But then one finds that the two electrons, despite the exclusion principle, are exactly in the same operator-state, therefore they are two identical and truly indistinguishable entities (and not a single entity, as Leibniz's principle would require). The same of course holds for bosons, in an even stronger way, as for them Pauli's principle does not even apply (de Ronde & Sassoli de Bianchi 2019).

Summing up, the true mystery of quantum indistinguishability lies in the fact that one can have composite systems, formed by truly indistinguishable entities, which are nevertheless able to remain a collectivity of individuals. We know that because some of their intrinsic properties are *extensive*. If we have a system of $N$ electrons, and we can measure its total electric charge, we will find it to be $Ne$, and not $e$. Hence, we know the system is truly formed by many entities, and not by a single entity. But when we observe their overall statistical behavior, we also know they are genuinely impossible to distinguish from one another, even in principle (unless one uses some specific means to force a distinction).

Thus, we are witnessing the possibility (which is an impossibility for spatiotemporal entities) of *being many and at the same time being genuinely indistinguishable*, which is the conundrum that quantum indistinguishability asks us to solve.

## 3 A new ontology and metaphysics

Having explained in some detail what the main quantum mysteries are, it is time to introduce the new ontology and metaphysics that can address such mysteries and transform quantum mechanics in an easy-to-understand theory, without taking away any of its depth and the amazement it produces in those who deeply reflect about its content.

To explain how this new ontology and metaphysics emerged, it is useful to briefly recall how quantum mechanics came into being. As is well known, in the beginning of the twentieth century the entire edifice of classical physics was shaken by several new experimental results, impossible to explain with the available physical theories, like the black-body's thermal radiation, the photoelectric effect, the atomic spectra, just to name the most important. To address these unexpected results, some brilliant scientists came up with a completely new theoretical edifice. Two lines were initially developed, one based on the *algebra of matrices*, initiated by Werner Heisenberg, and the other based on *differential equations*, initiated by Erwin Schrödinger. These two lines, although apparently very different, were made to converge by Paul Dirac (1930) and John von Neumann (1932), towards a very general unified scheme, using the mathematics of *Hilbert spaces* and *linear self-adjoint operators*.

These developments turned out to be extremely successful, but there was a flat. Despite the new formalism was highly predictive and extremely precise in accounting for all the new data collected in the laboratories, it remained unclear what it was all about, so much so that we are still here today, more than ninety years after the full enunciation of standard quantum mechanics, conjecturing about how it should be interpreted. In retrospect, we can say that part of the difficulty in elaborating the new Hilbertian quantum formalism, in the early days of its development, was to recognize that it was essentially a probabilistic theory, when equipped with the Born rule (Born 1926), deducible from Gleason's celebrated theorem (Gleason 1957), and that the quantum probabilistic model was very different from the classical one, axiomatized by Andrey Kolmogorov (1933) around the same years when von Neumann also proposed an axiomatization of quantum mechanics.

Mutatis mutandis, a similar situation happened in a very different domain of investigation: *psychology* and *economics*. Indeed, for quite some time theories involving the modeling of collections of human agents considered the latter to be entities behaving in a purely rational way. These were for instance the consumers or the companies in a free market, who were assumed to always maximize, with their decisions, all sorts of utilities. Some trace back this hypothesis to Aristotle himself, when he stressed that rationality is the crucial attribute differentiating human beings from animals. The problem is that with time this observation grew into the prejudice that humans would always be rational, when taking decisions, whereas the truth is that, although they certainly can be rational, when they try hard enough, they rarely are in practice.

This was strongly emphasized in the work of Amos Tversky and Daniel Kahneman, who explored numerous aspects of the psychology of *intuitive beliefs and choices* (Kahneman 2003) and the associated *bounded rationality* (Simon 1955). In other words, a large amount of data was accumulated over the years, describing all sorts of cognitive situations of human decision making, and similarly to the situation of physicists before the quantum revolution, theoretical psychologists, and economists, were also confronted with all sorts of results that appeared to be fundamentally irreconcilable with the classical rational-agent probabilistic models, based on Kolmogorovian probabilities and Boole's algebra, which is the algebra of logic and rationality.

In other words, as for quantum entities, whose behavior appeared to be impossible to understand, it was believed that one could not find any coherence in the irrational part of the human mind, governed by subconscious, instinctive, associative, intuitive, and similar processes. But things radically changed when some physicists, approximately two decades ago, had the idea of trying modeling *meaning entities* as if they were quantum entities. More precisely, the idea was to use the mathematical formalism of quantum theory as a *non-Kolmogorovian probabilistic model* to account, in a general and principles-based way, for all sorts of cognitive situations that were otherwise impossible to account for in a non-ad hoc way, when using classical probability models.

A new field emerged from these studies, today called *quantum cognition*, which saw the light in the nineties of the last century, thanks to the intuitions developed in our Brussels' group (Aerts & Aerts 1995, Aerts et al. 1999, Gabora and Aerts 2002, Aerts and Gabora 2005a,b) and thanks to the work of other initiators of this growing domain of investigation, like Andrei Khrennikov (1999) and Harald Atmanspacher et al. (2002). Since that time, impressive results were obtained in the modelling of humans' probability judgment errors, decision-making errors, and more generally in the representation of knowledge and meaning; see for instance Busemeyer & Bruza (2012) and the references cited therein. In parallel to that, similar insights were obtained in the related field of information retrieval (Aerts & Czachor 2004, Rijsbergen 2004, Widdows 2004), also addressing the problem of finding ways to model the level of meaning, particularly that contained in written documents, and in this case as well promising results were obtained using numerous notions derived from the quantum formalism; see for instance Melucci (2015) and the references cited therein.

It would bring us too far from the scope of this article to do the exegesis of the numerous insights and developments that have led to the modeling of all sorts of cognitive situations, using the different quantum notions and the power of quantum mathematics. What is worth emphasizing, however, is that there were two main modeling strategies. One, probably the most widespread today, was to use the quantum vector-states to describe the subjective beliefs of the persons responding to specific cognitive situations. The other, which since the beginning was adopted in our group, and which we think is, in a way, more fundamental, is to use the quantum states to describe the *modes of being* of the conceptual entities themselves, to be understood as entities conveying specific meanings (be them actual or potential).

Consequently, human agents participating in cognitive tests are viewed in our approach as playing the role of *contexts* for the different conceptual entities, sensitive to the meaning they convey, and depending on the design of the experiments, these contexts can either be deterministic or indeterministic. This means that subjective beliefs are considered to be part of the *meaning driven interactions* arising between the conceptual situations and the human participants reacting and deciding on these situations (Aerts et al. 2016, 2018a).

Now, this unexpected discovery that quantum mechanics led itself extremely well to the construction of a qualitative and quantitative theory of human concepts and their combinations, and of their interactions with the human minds, brought one of the authors to a bold hypothesis, which was initially formulated in the form of a question (Aerts 2009):

*If quantum mechanics, as a formalism, models human concepts so well, perhaps this indicates that quantum entities themselves are conceptual entities?*

Behind this question, there was the idea that maybe it is not a coincidence that the mathematical formalism of quantum mechanics is so well-appointed to describe numerous aspects of the human cognitive activity. This could be the case precisely because of the nature of the microphysical entities. More precisely, the conceptuality interpretation introduces the hypothesis that there is a (possibly fundamental) duality in our reality: on one hand, there are the symbolic entities forming languages, the so-called *concepts*, which through their combinations are able to carry meaning and also create new meaning; and on the other hand, there are the *cognitive entities*, the minds, sensitive to the level of meaning, which communicate by interacting with the conceptual entities forming their languages.

In other words, the conceptuality interpretation posits that (Aerts 2009, 2010a,b, 2013, 2014, Aerts & Sassoli de Bianchi 2018, Aerts et al. 2020, Aerts & Beltran 2020, Aerts et al. 2024b,c):

(1) the microphysical entities are conceptual in nature, whereas the entities made of ordinary matter are cognitive in nature;
(2) the mind-language duality of the human cultural domain mirrors the fermion-boson duality of the physical domain, in the sense that bosons and fermions are entities belonging to a more ancient cultural domain, where bosons are the natural building blocks of languages and fermions are the natural building blocks of the cognitive structures.

In other words, the new ontology says that there are two fundamental kinds of entities out there: the conceptual ones, carrying meaning (and bosons would be their archetype) and the cognitive ones, sensitive to meaning (and fermions, when they form large aggregates would be their archetype). Therefore, the new metaphysics says that the nature of our physical reality is that of a *duality of concepts and minds* which have evolved symbiotically. If this is correct, then, quoting from (Aerts & Sassoli de Bianchi 2018):

> [...] something similar to what happened in our human macro-world, with individuals using concepts and their combinations to communicate, may have already occurred, and continue to occur, mutatis mutandis, in the micro-realm, with the entities made of ordinary matter communicating and co-evolving thanks to a communication that uses a language made of concepts and combinations of concepts that are precisely the quantum entities and their combinations.

Because of the above, our universe would be fundamentally *pancognitivist*, in the sense that everything in it would *participate in cognition*, with human cognition being just a very recent episode of it, expressed at a very specific organizational level.

## 4 Explaining the quantum mysteries

In Section 2, we have identified five main quantum mysteries. In this section, we address them one by one, by using the conceptuality interpretation, i.e., by adopting the ontology and metaphysics we introduced in the previous section.

### 4.1 Explaining quantum superposition

In Section 2.1, we stated that the real mystery of quantum superposition manifests when the superposition is about mutually exclusive spatial properties, as in the case of two states expressing the localization of a micro-entity in two separate and distant regions of space. Since such entity cannot be consistently described as something widespread in space, the superposition state necessarily describes a situation of unactualized spatial properties, i.e., a *non-spatial state*. The question then arises: "What does it mean to be in a non-spatial state?" The answer is: "It means that we have to refrain forcing an *object-view* on a quantum entity." In a nutshell, a quantum micro-entity, if it can be in a non-spatial state, it is because it is not a particle, it is not a wave, and it is not a field (Sassoli de Bianchi 2013). Quoting Jean-Marc Lévy-Leblond (2018):

> It is to be realized today [...] that quantum theory does exist and that its concepts, after a century of collective practice, are deeply rooted in the present common sense of working physicists. These concepts need no longer be approached from classical ones, but may, and should, be taken at their face value. Such an intrinsically quantum understanding leads one to recognize that the objects of quantum physics are not either waves or particles, as duality would want us to believe; they are neither waves, nor particles, even though they do exhibit, under very particular circumstances, two types of limit behavior as (classical) waves, or (classical) particles [...]. It has been proposed to stress this ontological point by calling them 'quantons'.

The idea of using a new term to designate the microscopic quantum entities is certainly advisable, to avoid the confusion of associating them with improper corpuscular-like or wave-like spatial notions, or even with the notion of a field having well-defined spatial properties. But the fact remains that using a new term does not help to better understand what the newly baptized *quanton* truly is. The conceptuality interpretation, on the other hand, provides a simple and direct answer: *a quanton is a conceptual entity*. Not a human concept, but an entity sharing with human concepts a similar *conceptual nature*, in the same way an acoustic wave shares with an electromagnetic wave a similar *undulatory nature*, while remaining very different entities.

If the above is true, then the mystery of non-spatiality disappears, in the same way that there is no mystery in observing that, say, the Italian language, or the Dutch language (the two mother tongues spoken by the authors), are non-spatial entities. Try asking yourself where they are located. One might be tempted to answer that they are like spatially extended entities, which can be found in all those places like Italian or Dutch books and journals, Italian or Dutch movies, and where people are speaking them to communicate. This, however, would be a wrong answer. Indeed, it confuses an entity with the traces it can leave in the different substrata of our spatiotemporal theater. This example of languages, as entities whose non-spatial nature is evident to everyone, also shows the complexity such entities can give rise to. Very often indeed, a third language, like English, is used when two people who lend themselves to two different languages, such as Italian and Dutch, need to communicate, as the authors can attest.

To clarify this fundamental point, consider the double-slit experiment (see Figure 1). When a photon leaves a trace on the final detection screen, such trace is not to be mistaken for the photon itself, which by the way usually no longer exists once it has been absorbed by the screen. Seemingly, there is an aspect of a *language*, as a *conceptual entity*, that is purely abstract, hence non-spatial, which is different from all the traces that such abstract entity can leave on the different available supports, including the brains and minds of the humans who use it. We can easily understand that this abstract aspect of a language, not contained in its possible spatializations, is what we usually denote *meaning*. For example, the communication between a native speaker of Italian and a native speaker of Dutch, in English, uses three different languages, of which only the English one is mastered by both speakers, but the meaning of their communication, carried by the English spatialization, belongs to this more abstract aspect of human language, which is almost totally independent of it being expressed in a specific language.

Now, the most scientifically sound theory on the evolutionary origins of human language starts from the hypothesis that sign language existed first, and that spoken language emerged from the slow adoption of gestures previously performed with hands and other parts of the body, by the inside of the mouth and throat, the larynx, with the emission of sounds in which the *meaning* originally *spatialized* and *encoded* in gestures was then transferred to another spatial carrier, that of sound waves. Scripture, an additional and different spatial medium, came later, but still with the same original purpose of exchanging meaning between human minds.

A word is generally regarded as already belonging to a specific language, and in this sense subject to an initial form of spatialization, even if a space in the strict sense is not yet involved at this stage. However, when we analyze, in Section 4.4, how in a language the connecting element "and" of conjunction behaves, unlike the connecting element "or" of disjunction, we will see that the formation of a primitive spatial structure is in fact already present. We use the notion of *concept* here, distinguishing it from the notion of *word*, when we refer to an entity that is a carrier of meaning independently of the words in different languages that also carry such meaning. Concepts, in this sense, are to be understood as the building blocks of the most abstract level of human language. For example, in addition to the words *arbitrario*, in Italian, *willekeurig*, in Dutch, and *arbitrary*, in English, there is likewise the *concept of arbitrariness*, which carries the meaning that each of the words mentioned above carries, but independently of the spatialization promoted by a specific language. Then, further possible spatializations consist of the transition from a word in a given language to, for example, the oral expression of such word, or its writing on a given material medium. And it is in this latter process that the human conceptual world "touches" the material conceptual world, as we also explain in Section 4.4.

A peculiarity is that there is a great number of languages whose spatializations, whether in spoken or written form, are all very different from each other, so that a person who masters the spatialization of one of them, the so-called mother tongue, will be able to master the other spatializations only through long learning, and never spontaneously, even though all these different languages convey essentially the same meaning. A hypothesis that seeks to explain this gradual transfer of the body's original gestures to the inside of the mouth, holds that language was also used to communicate more discreetly and secretly, as a means of discrimination. The dramatic story of the Tower of Babel, in the Bible, is perhaps a reflection of this period when language became an instrument of separation between different groups of the human species, who became mutually incomprehensible. This evolutionary-historical reflection on human language is particularly interesting for the conceptuality interpretation, whose main motivation lies, precisely, in trying to understand the reasons for our inability to comprehend quantum mechanics. What we have mentioned about human language may indeed alert us to the fact that, just as we do not master the spatializations of the languages we do not speak, it is mainly with the spatialized elements of the quantum reality that we are experimentally confronted. Therefore, we can assume that this spatialization of it, corresponding to a language that is non-native for us, contains a part of specificity that, somehow, veils the *meaning* that the pieces of fermionic matter communicate with each other, through quantum language.

Having said that, still on the subject of the crucial distinction between a conceptual entity carrying meaning and the traces it can leave on different typologies of spatiotemporal supports, in Aerts et al. (2018a) we wrote the following, in relation to the corpora of written documents forming the World Wide Web:

> [...] we firstly have to make clear the distinction between two kinds of Web: the standard (spatial) Web, made of actual webpages, formed by specific collections of letters and words, printed on paper or encoded in computers' memories, and the 'meaning entity' that we can associate with it, formed by concepts existing in different combinations, which is the object of our modeling. This (non-spatial) meaning/conceptual entity, which we will simply call the QWeb (i.e., the 'Quantum Web'), is of course intimately related to the standard Web, in the same way that a concept, say the concept *Fruits*, is intimately related to the different possible printed words that can be used to indicate it.

So, if a quantum entity is a conceptual entity, carrying meaning, i.e., an entity of the same nature of a human concept, which is very different from a printed word of that concept, or a full text written to explain its meaning (think of the text explaining a given word in a vocabulary), it becomes very clear, if not self-evident, why it cannot be a spatial entity. Consider the difference between these two sentences: "In this moment Massimiliano is visiting Diederik in Brussels," and "In this moment Massimiliano is either visiting Diederik in Brussels or he is in his home in Lugano." If the first sentence describes a Brussels' localized state of the conceptual entity *Massimiliano* (not to be confused with Massimiliano's body), hence, not a spatial superposition state, the second one does describe a superposition state of two states of *Massimiliano* that are spatialized. And of course, in the conceptual realm, these states present no mystery. Hence, if quantum entities are conceptual in nature, neither their non-spatiality nor their superposition states should present any mystery. Of course, they still come as a surprise, given the historical prejudices that lead us to consider as existing only what exists as an object, in the sense of an entity permanently present in space and time.

The above explanation becomes much more convincing when one shows that interference effects, resulting from the existence of superposition states, like in typical double-slit experiments, can easily be explained using the conceptuality interpretation (Aerts et al. 2020, Sassoli de Bianchi & Sassoli de Bianchi, 2025), also because similar effects can also be evidenced in the human conceptual domain.

Consider for instance a study by James Hampton (1988), where participants were subjected to the following 24 exemplars of fruits and vegetables (see Figure 7): (1) *Almond,* (2) *Acorn,* (3) *Peanut,* (4) *Olive,* (5) *Coconut,* (6) *Raisin,* (7) *Elderberry,* (8) *Apple,* (9) *Mustard,* (10) *Wheat,* (11) *Ginger root,* (12) *Chili pepper,* (13) *Garlic,* (14) *Mushroom,* (15) *Watercress,* (16) *Lentils,* (17) *Green pepper,* (18) *Yam,* (19) *Tomato,* (20) *Pumpkin,* (21) *Broccoli,* (22) *Rice,* (23) *Parsley,* (24) *Black pepper*. More precisely, they were confronted with three interrogative (measurement-like) situations. They were asked, with regard to the above set of exemplars, to:

(1) to choose a typical exemplar of *Fruit*;
(2) to choose a typical exemplar of *Vegetable*;
(3) to choose a typical exemplar of *Fruit or vegetable*.

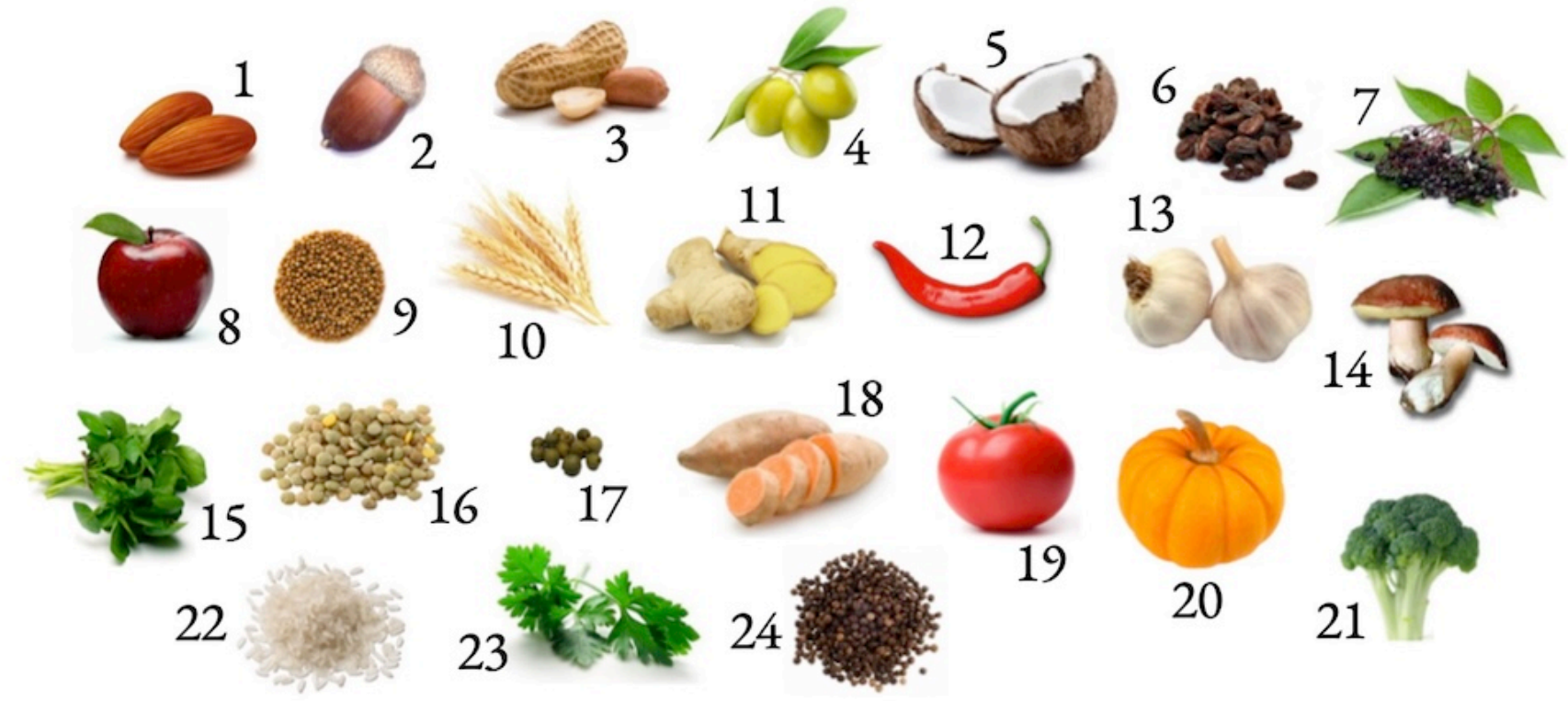

**Figure 7** (color online). A depiction of the 24 exemplars of fruits and vegetables used in Hampton's 1988 study.

In these three different situations, the relative frequencies with which the different exemplars were selected were calculated, corresponding to the experimental probabilities that a human subject, confronted with these three situations, would choose those specific exemplars. Without going into details, let us just mention that the values of the obtained probabilities were impossible to understand based on pure classical reasonings. For example, the third situation could not be understood as being some kind of average of the first two situations. This because the probabilities for some of the exemplars were too strongly *overextended* with respect to the expected average values, whereas others were too strongly *underextended*.

These cognitive-conceptual effects of overextension and underextension are what in physics one usually describes as *constructive interference effects* and *destructive interference effects*. In other words, interference effects occur when concepts are combined (here the combination of the concept *Fruit* with the concept *Vegetable* in the disjunctive sentence *Fruit or vegetable*), as conceptual combinations are like superposition states. This can be highlighted in a very convincing way when the overextensions and underextensions are modeled and analyzed using the Hilbertian formalism of quantum mechanics. In the present case, the description is equivalent to that of a double-slit experiment, with situation (1) corresponding to the case where only the first slit is open (the *Fruit*-slit), situation (2) to the case where only the second slit is open (the *Vegetable*-slit), and situation (3) to the case where both slits are open (Aerts, 2009, Aerts & Sassoli de Bianchi 2017b).

More specifically, using two-dimensional wave-functions able to correctly predict (via the Born rule) the outcome probabilities of situations (1) and (2), then constructing a suitable normalized superposition of them, also able to predict the probabilities of situation (3), a complex interference pattern is revealed, reminiscent of those obtained in the phenomena of *birefringence*; see Figure 8.

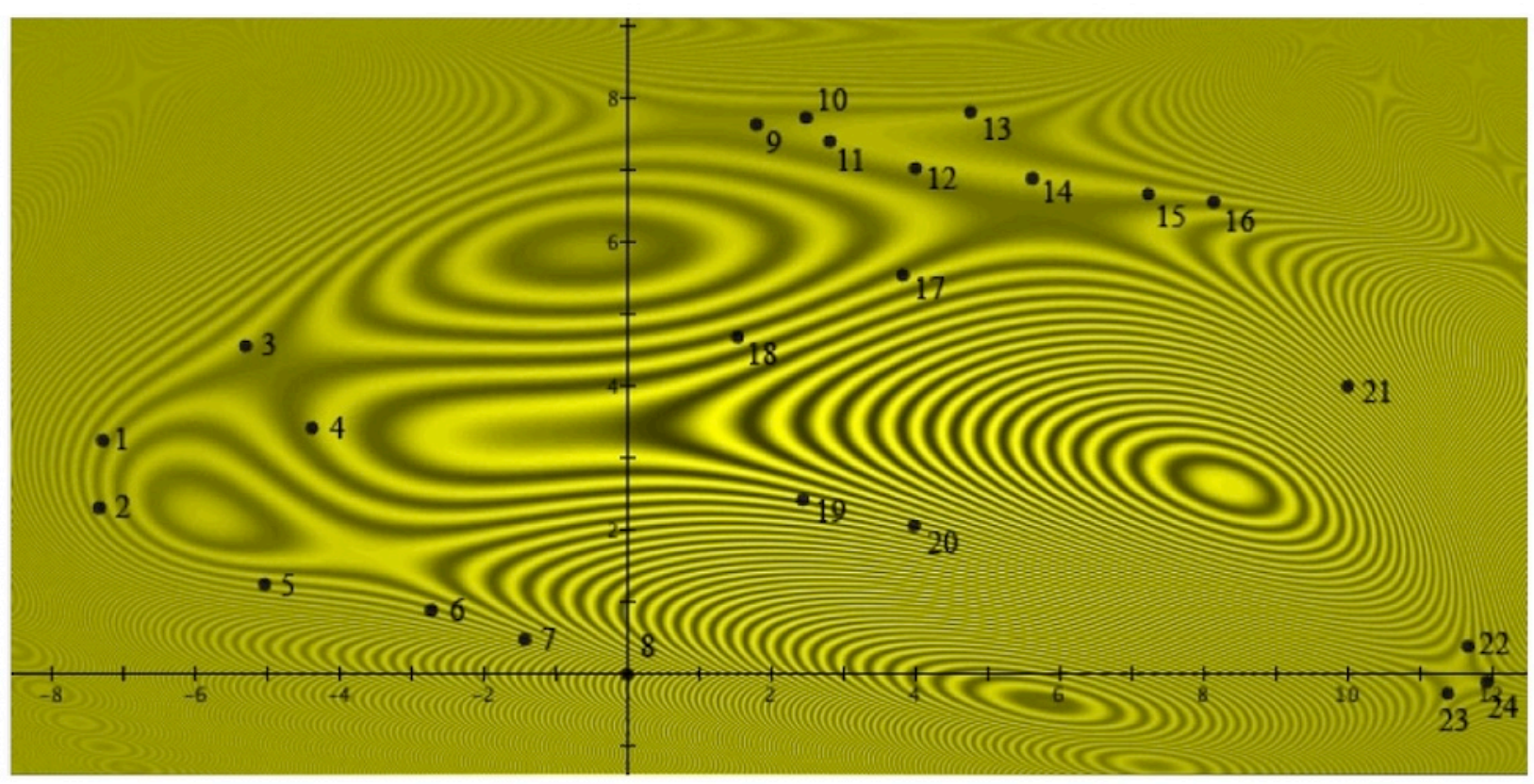

**Figure 8** (color online). The effects of overextension and underextension of the probabilities observed by Hampton (1988), in the situation (3), where participants were asked to choose a typical exemplar of *Fruit or vegetable*, when modeled using the quantum formalism give rise to a particular interference pattern on the two-dimensional *screen of exemplars*. Data were here fitted in a two-dimensional wave function $\psi_{(3)}(x,y) = [\psi_{(1)}(x,y) + \psi_{(2)}(x,y)]/\sqrt{2}$, obtained by superposing two Gaussians, $\psi_{(1)}(x,y)$ and $\psi_{(2)}(x,y)$, modeling the data associated with situations (1) and (2), respectively. Outcome probabilities were obtained via the Born rule, i.e., as the square modulus of the wave function, $|\psi_{(3)}(x,y)|^2$. In the figure, the brightest regions are those of highest probability (Aerts, 2009).

## 4.2 Explaining quantum measurement

In Section 2.2, we observed that a *quantum measurement* is like a weighted symmetry breaking process, and that its mystery is in part related to the fact that, on the one hand, it requires the existence of non-spatial superposition states, and on the other hand, it demands to understand the nature of the uncontrollable fluctuations that would bring such states into spatial ones (in case the measurements are about spatial localization properties), and there is also the question of understanding why all this would be governed by the Born rule.

We have already considered the issue of superposition and non-spatiality in the previous section. Concerning the second issue, the origin of the Born rule, here as well the conceptuality interpretation offers a very clear answer. Indeed, within its paradigm, a measuring instrument being a macroscopic material object mostly made of fermions, it is understood as behaving as a cognitive entity. Hence, a quantum measuring process is to be interpreted as a *cognitive interrogative process* and should be explained by referring to what typically happens during a process of this kind.

Here of course one needs to distinguish the experimenter's interrogative process from the interrogative process of which the measuring apparatus as such would be part of. It is easy to confuse these two cognitive layers. The first one is the question the experimenter asks, when conceiving and carrying out a given experiment. The outcomes s/he obtains are of course the answers to her/his query. This is not, however, what the conceptuality interpretation refers to, when it considers a quantum measurement as a cognitive interrogative process. More precisely, in a quantum measurement performed in a laboratory there are two cognitive entities: the experimenter and the apparatus. Both undergo interrogative processes, but they are not the same, although they are certainly related. The experimenter, in a sense, uses the measuring apparatus to answer her/his experimental question, which s/he formulates using the human language. The answer s/he obtains is also formulated via the human language. For instance, in a given experiment the question might be (we are obviously simplifying): "The photon emitted by the source, where is it located?" And when a spot (a trace) appears on the detection screen, the experimenter gets the answer that: "the photon (before being possibly destroyed) was located where the spot appeared."

On the other hand, the detector screen, understood as a cognitive-like entity able to understand the meaning carried by the photon, answers a different question, which cannot be formulated in our human language. When we use the quantum formalism, in a sense we are closer to the language that the detector screen understands, which is carried by the photon, but certainly it remains a translation, into the human symbolic realm, of something that is extremely distant from our human culture. This does not mean, however, that we cannot try to approximate the meaning carried by such photon-screen communication; see for instance how the double-slit experiment was "translated" into our human language in Aerts et al. (2020), to deduce the (fringe) pattern of answers that is typically obtained. Here we must always keep in mind the Italian saying "traduttore, traditore," i.e., "translator, traitor," and that we humans cannot even communicate with an ant. But we can certainly understand that ants do communicate, and even if we do not understand the specifics of their language, we can certainly observe the pragmatic effects of their communication.

The conceptuality interpretation posits that we can do the same when observing the result of the interaction between microscopic entities and measuring apparatuses. We do not need to understand the (non-human) meaning they are exactly exchanging, but we can nevertheless understand that they are indeed exchanging meaning. This is in fact the central hypothesis behind the conceptuality interpretation: that the quantum formalism is a formalism that we humans have developed to model, without knowing it, meaning entities, and this is the reason why it turned out to be so successful in also modeling the human cognitive domain.

Coming back to the quantum measurement process, a measurement apparatus is like a mind-entity who is forced to answer a given question. Now, there are special situations where the measured entity is in an eigenstate of the measurement. This corresponds to a cognitive situation where the answer to the posed question is predetermined. Like when a person is asked if s/he likes pizza or if s/he likes waffles. The answer to both questions (apart anomalies) exist even before the questions are formulated and addressed. When, instead, the measured entity is in a superposition state, it corresponds to the situation where a person is asked a question which requires to "take a position" on the spot. Like when someone who has never tasted a given dish, but only saw it once in a picture, is asked if s/he would enjoy eating it. Here we can easily understand that the answer is not predetermined, and that even the person answering the question cannot predict in advance what her/his answer will be. Based on partial information, s/he will have to conceptualize the situation in more abstract terms, then guess a possible concrete answer, i.e., actualize one of the possible responses.

The reason why this is relevant for understanding the emergence of the quantum mechanical Born rule is that one can easily understand that when a person is forced to take a position, by answering a question s/he never reflected about before (so s/he does not have all the elements to know in advance what the answer will be), when picking a possible answer all sorts of *intrapsychical* and *extrapsychical* fluctuations will be involved in such a process of (weighted) symmetry breaking, during which only one among the possible (potential) answers will be actualized.

More precisely, one can distinguish, in the response process, two stages. In a first stage, the person brings the meaning of the situation as close as possible to the meaning of the different possible answers. This builds a *maximally unstable state of mental equilibrium*, i.e., a maximally unstable *tension* formed by the different competing meaning-connections with the different possible answers. At this point comes the second stage: the smallest fluctuations, i.e., the smallest disturbances in the evaluation process, will be able to break the equilibrium (to *reduce* the tension) in favor of one of the competing connections, producing in this way a genuinely unpredictable outcome.

Why the above description of a mind process (not to be confused with the correlated brain process) would be relevant to explain what goes on behind the spatiotemporal scenes of a quantum measurement? Because quantum probabilities can in fact be explained in terms of fluctuations that are actualized at the level of the interactions between the measurement apparatus and the measured entities, called *hidden-measurements interactions* (Aerts 1986, Aerts & Sassoli de Bianchi 2014, 2017b). In other words, a measuring apparatus, behaving like a cognitive entity, requires the existence of competing answers, hence of a growing *tensional equilibrium*, which in turn requires the possibility of breaking the symmetry of that equilibrium, hence the existence of a symmetry breaking process, which in turn requires the presence of fluctuations (Aerts & Sassoli de Bianchi, 2015b,c).

But there is more. When the standard quantum formalism, which only uses (pure) *vector-states* to represent genuine states, is extended into a Blochean formalism, where also *operator-states* (density operators) can represent genuine states, one can precisely describe the two-stage process we mentioned above. Indeed, in such an extended formalism, which constitutes a natural completion of the standard quantum formalism, specific structures appear, precisely describing the *abstract potentiality regions* at the origin of the fluctuations and of the collapses. And one can then explain a measurement process as a two-stage process (actually, a three-stage process, when degenerate outcomes are also considered), where in the first stage the point representing the state within the generalized Bloch sphere approaches the structure defining the tensional equilibrium, producing a so-called *decoherence* of the state, and in the subsequent stage, when the equilibrium is broken, an irreversible process selects one among the possible outcomes. The beauty of this is that the (geometric) structures in question, which are *simplexes* inscribed in the *generalized Bloch sphere*, allow to exactly predict the probabilities of the Born rule, which therefore can be derived in a non-circular way (Aerts & Sassoli de Bianchi 2014, 2017b); see Figures 9 and 10.

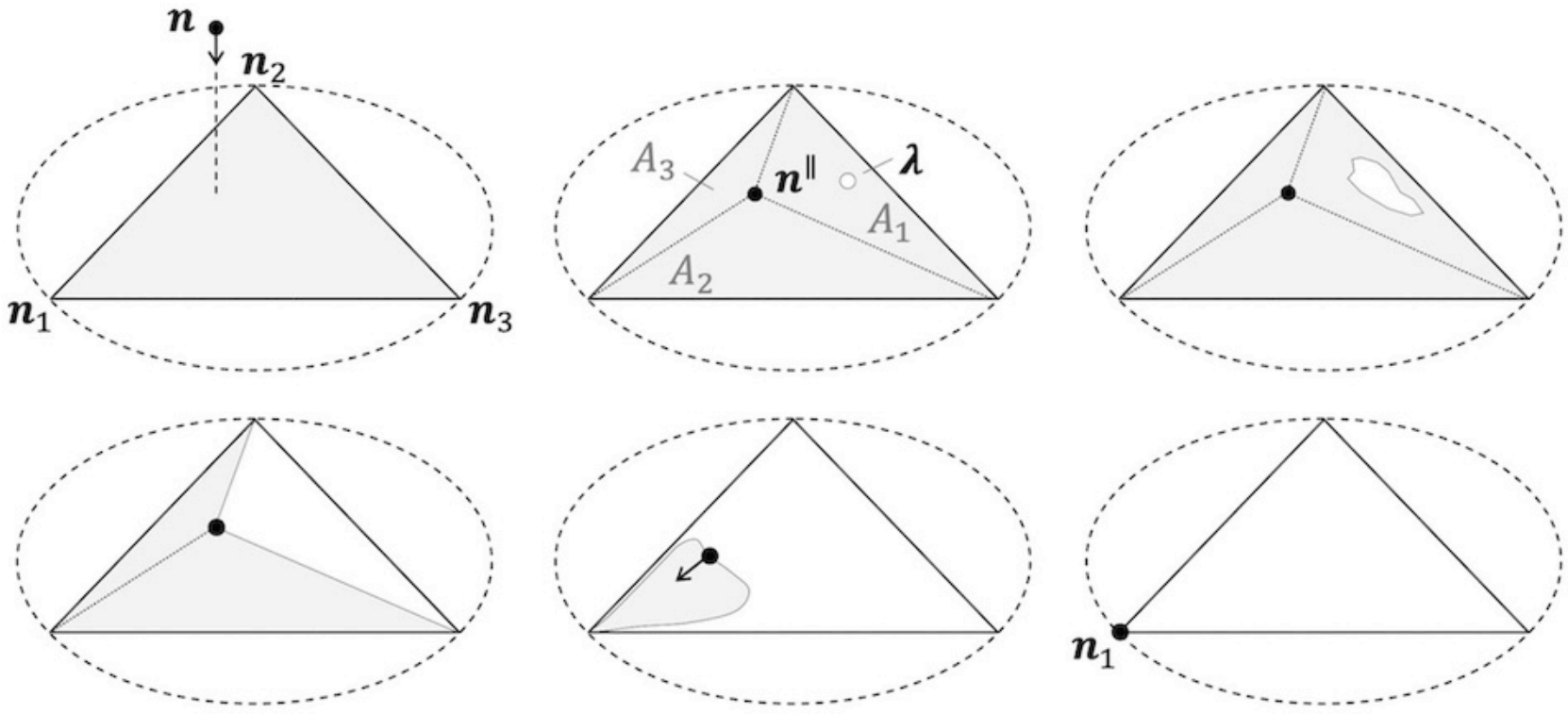

**Figure 9** The unfolding of a quantum measurement process, here with three distinguishable outcomes, in the extended Bloch representation. The point particle representative of the state, initially located in position $\boldsymbol{n}$ at the surface of the (here 8-dimensional) Bloch sphere, approaches the potentiality region (decoherence process), which can be described as a triangular elastic membrane. When it reaches the on-membrane point $\boldsymbol{n}^{\parallel}$, a maximally unstable state of equilibrium is created, characterized by tension lines between the different competing answers: $\boldsymbol{n}_1$, $\boldsymbol{n}_2$ and $\boldsymbol{n}_3$. These tension lines define three distinct subregions, $A_1$, $A_2$ and $A_3$, the relative areas of which are given by the Born rule. When fluctuations break the equilibrium, this corresponds to a disintegration of the membrane at some unpredictable point $\boldsymbol{\lambda}$, producing the complete collapse of the associated subregion, here $A_1$, causing it to lose two of its anchor points, drawing in this way the point particle to its final location, here $\boldsymbol{n}_1$.

It is not the purpose of the present article, which wants to remain accessible to a wide multidisciplinary audience of scientists and philosophers, to enter into the mathematical details of the above depiction. It is however important to emphasize that what we just outlined in qualitative terms can be expressed in a very precise and powerful mathematical language. Hence, although the hidden-measurements solution to the measurement problem was historically advanced starting from considerations that are not directly related to the conceptuality interpretation, in retrospect we can say that the latter contains within it all the key elements able to lead to such solution.

To further strengthen the last statement, we also take the opportunity to say a few words about the notion of *quantization*, here to be understood in the sense historically attributed to it by Max Planck, in the article that initiated the Old Quantum Theory (Planck, 1900). Indeed, quantization can also be understood by reasoning about the workings of human perception (Aerts & Beltran 2022, Aerts & Aerts Arguëlles 2022). Note that when the quantum structure of human concepts was initially identified (Aerts & Gabora 2005a,b), it was not yet realized that in all forms of human perception there exists a phenomenon which can be related to quantization, consisting of a systematic *warping of stimuli* by human perception, called *categorical perception* (Goldstone & Hendrickson 2010). The insight about categorical perception, extensively studied by cognitive scientists, came to us only in recent times, and the reason of this delay is the partial lack of communication between two subgroups of cognitive scientists: on the one hand, that which studies human concepts, and on the other hand, that which studies human perception.

The warping effect of categorical perception is apparently valued only by the latter subgroup, as a crucial mechanism establishing what human categories or concepts actually are. Now, in the years when we developed the quantum modeling of human concepts (Aerts & Gabora 2005a,b), we relied mainly on the theoretical and experimental studies by the concept subgroup (Smith & Medin 1981), where the warping effect of categorical perception was not taken in consideration. Recently, however, we realized that this mechanism could explain the emergence of quanta, providing further evidence for the research domain of *quantum cognition* and adding to the explanatory power of the conceptuality interpretation.

But what is this phenomenon of categorical perception, and why does it contain a possible explanation for quantization? As we said, it corresponds to a *warping* that occurs in all forms of human perception between, on the one hand, what psychologists call the *stimuli* and, on the other hand, the so-called *percepts*, by which they designate *that which is experienced* by the person who perceives. The warping consists of perceiving stimuli that fall within a same category as more similar, and stimuli that fall into different categories as more different, where "more similar" and "more different" are defined with respect to how similar or different the stimuli are with respect to their most possible objective observation, and how similar or different the percepts are with respect to the perceiver's experience. In other words, the warping causes some stimuli to become perceptually more distant, and others to become closer, forming *quanta*, i.e., giving rise to a phenomenon of clumping and discretization.

Let us illustrate the phenomenon with the example of the perception of colors on the side of the perceiver, and the physical phenomenon of light, characterized by frequencies, on the side of the stimuli [see Aerts Arguëlles (2024) for a detailed analysis of this example]. People see colors as shades of about eight basic colors, *red*, *green*, *yellow*, *blue*, *purple*, *pink*, *orange* and *gray*. At least that's what the research shows, but it is also known that the number of colors that can be distinguished depends on the human cultures and can vary from two to eight (Brent & Kay 1969). Some of these shades are considered to be the best example of their color, called *focal colors* (Brent & Kay 1969), or *prototype colors* (Rosch 1973), and correspond to well-defined frequencies of the electromagnetic radiation.

The warping of categorical perception at work in the case of colors means that two different frequencies that are both perceived as, say, green, hence as shades of one and the same color, are perceived as more similar as compared to two different frequencies of which one is perceived as green and the other as, say, yellow, hence as shades of two different colors, even if the difference in frequency of both pairs is the same when measured in a physics' laboratory. Note that the warping is a *contraction* if the percepts are shades of the same color, and a *dilation* if the percepts are shades of two different colors.

This categorical perception mechanism can also be found in quantum measurements, in the sense that it would be the decoherence and collapse of a pre-measurement state what produces the phenomenon of categorical perception. Here we will just give a brief hint of how this works, using a simple but significant example, and refer to Arguëlles (2024) for details. It will also be an opportunity for us to explain how Bloch's extended representation works in the case of a two-dimensional entity, namely a qubit.

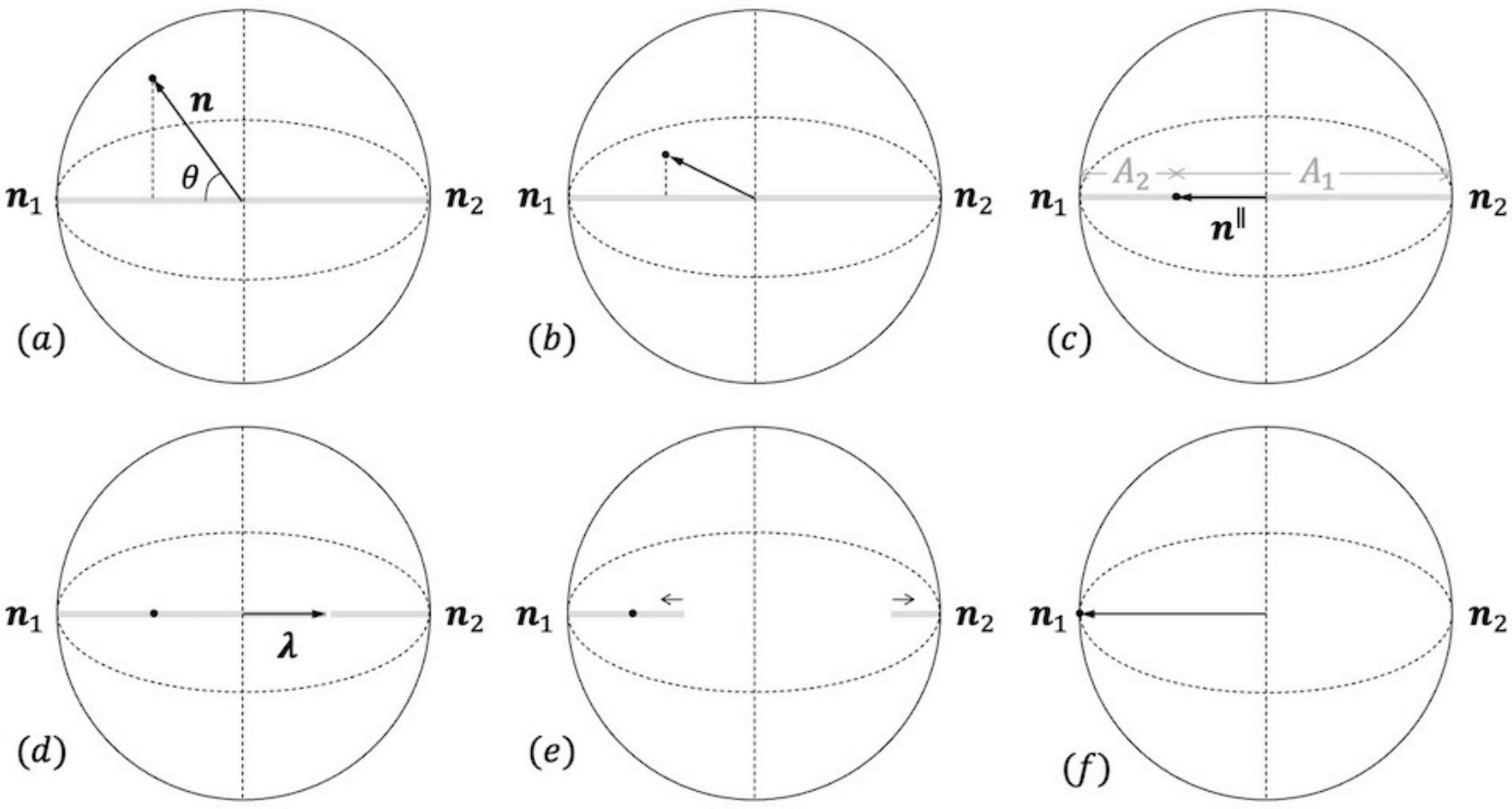

**Figure 10** The unfolding of a quantum measurement process, here with only two outcomes. The point particle representative of the state, initially located in position $\boldsymbol{n}$ at the surface of the 3-dimensional Bloch sphere, approaches the potentiality region (decoherence process) generated by the two outcome states $\boldsymbol{n}_1$ and $\boldsymbol{n}_2$. Such region can be described as an abstract one-dimensional elastic band (1-simplex). When the particle reaches the point $\boldsymbol{n}^{\parallel}$ on the band, it defines two distinct intervals, $A_1$ and $A_2$, the relative length of which are given by the Born rule. When the elastic band breaks, at some unpredictable point $\boldsymbol{\lambda}$, here assumed to belong to $A_1$, its contraction draws the point-particle toward position $\boldsymbol{n}_1$, corresponding to the collapsed outcome state.

To begin, we must determine what the stimuli and percepts are in the ambit of a quantum measurement. With the conceptuality interpretation in mind, we can treat a measurement as a cognitive process, hence the stimuli are represented by the vector-states (pure states), whereas the percepts, seen as the "perceived images" of these pure states, are represented by the corresponding operator-states (density operators). Now, as mentioned already, in the extended Bloch modeling of a quantum measurement (Aerts 1985, Aerts & Sassoli de Bianchi 2014, 2016), operator-states are obtained when one allows the one-dimensional projection operator describing a vector-state to *decohere*, i.e., to lose its non-diagonal matrix elements, when plunging into the Bloch sphere to reach the simplex generated by the outcome states. The process is described in Figure 9 for the 3-outcome case. Note that the generalized Bloch sphere is $(N^2 - 1)$-dimensional for a $N$-outcome measurement, hence, in the $N = 2$, it is 3-dimensional, and the measurement process can be fully visualized; see Figure 10.

To express that stimuli are warped into percepts, we need to measure the difference between two stimuli and between the two percepts corresponding to these stimuli. There is no obvious way to do so. At first sight, one might think that it would be sufficient to identify a *metric* on the set of quantum states, which could then be used to measure the difference between both stimuli (vector-states) and percepts (operator-states), but the problem is that there is no single natural metric on the set of all quantum states: many different metrics exist, all starting from a different vantage point on the collection of quantum states. One of the most important, called the *trace distance*, equals for qubit states to half the Euclidean distance in the Bloch sphere. This is a rather natural choice for measuring the distance between operator-states, but not for vector-states. Indeed, if $\psi_A$ and $\psi_B$ are two (two-dimensional, complex) unit vector-states, described in the Bloch sphere by two (three-dimensional, real) unit vectors $\boldsymbol{n}_A$ and $\boldsymbol{n}_B$, as per the formulae

$$|\psi_A\rangle\langle\psi_A| = \frac{1}{2}(\mathbb{I} + \boldsymbol{n}_A \cdot \boldsymbol{\sigma}) \qquad |\psi_B\rangle\langle\psi_B| = \frac{1}{2}(\mathbb{I} + \boldsymbol{n}_B \cdot \boldsymbol{\sigma})$$

where $\boldsymbol{\sigma}$ is a vector whose components are the three Pauli's matrices and $\mathbb{I}$ is the identity matrix (Aerts & Sassoli de Bianchi 2014), then a straight line going from $\boldsymbol{n}_A$ to $\boldsymbol{n}_B$, along which the trace distance would be determined, necessarily passes through points inside the sphere, representing operator-states. However, a natural notion of distance between pure states should only account for the vector-states, i.e., the pure states, that are in-between $\boldsymbol{n}_A$ and $\boldsymbol{n}_B$, i.e., a distance measured only on the surface of the Bloch sphere, corresponding to the length of the circular arc connecting them. More precisely, if $\theta_A$ and $\theta_B$ are the polar angles of $\boldsymbol{n}_A$ and $\boldsymbol{n}_B$ (see Figure 11), assuming here for simplicity that they have the same azimuthal angle, their vector-state distance, normalized to 1, is:

$$d_{pure}(\boldsymbol{n}_A, \boldsymbol{n}_B) = \frac{1}{\pi}|\theta_B - \theta_A|$$

Note that this distance also results from the angle between vector-states calculated from their Hilbert space inner product, which shows that it is a natural distance associated with the Bloch representation of pure states. More precisely, for pure states the *Fubini-Study distance* (Fubini 1904), which is a natural geometric metric on the complex projective Hilbert space, is given by:

$$d_{FS}(\boldsymbol{n}_A, \boldsymbol{n}_B) = cos^{-1}|\langle\psi_A|\psi_B\rangle| = cos^{-1}\sqrt{\frac{1}{2}(1 + \boldsymbol{n}_A \cdot \boldsymbol{n}_B)}$$

If $\boldsymbol{n}_A$ and $\boldsymbol{n}_B$ have the same azimuthal angle (see Figure 11), then $\boldsymbol{n}_A \cdot \boldsymbol{n}_B = cos(\theta_B - \theta_A)$. Since

$$\frac{1}{2}[1 + cos(\theta_B - \theta_A)] = cos^2\left(\frac{\theta_B - \theta_A}{2}\right)$$

and $\frac{1}{2}(\theta_B - \theta_A)$ is in the range $\left[-\frac{\pi}{2}, \frac{\pi}{2}\right]$, where the cosine is non-negative, we obtain: $d_{pure}(\boldsymbol{n}_A, \boldsymbol{n}_B) = \frac{2}{\pi} d_{FS}(\boldsymbol{n}_A, \boldsymbol{n}_B)$. Hence, the normalized distance measured on the surface of the Bloch sphere is directly proportional to the distance obtained using the Fubini-Study metric, also known as the *Bures metric*. Since the latter can be expressed in terms of the notion of *fidelity*, it also follows that when we consider angles, or equivalently circular arcs, as a measure of the natural distance between vector-states in the Bloch model, we are indirectly using the notion of fidelity to evaluate differences between stimuli.

Now, to analyze how the presence of the warping associated with categorical perception can also shed light on how to adequately map differences between quantum states, which is an important problem in quantum information theory, this would take us too far from the topic of this article, but it is our intention to continue this analysis in future work.

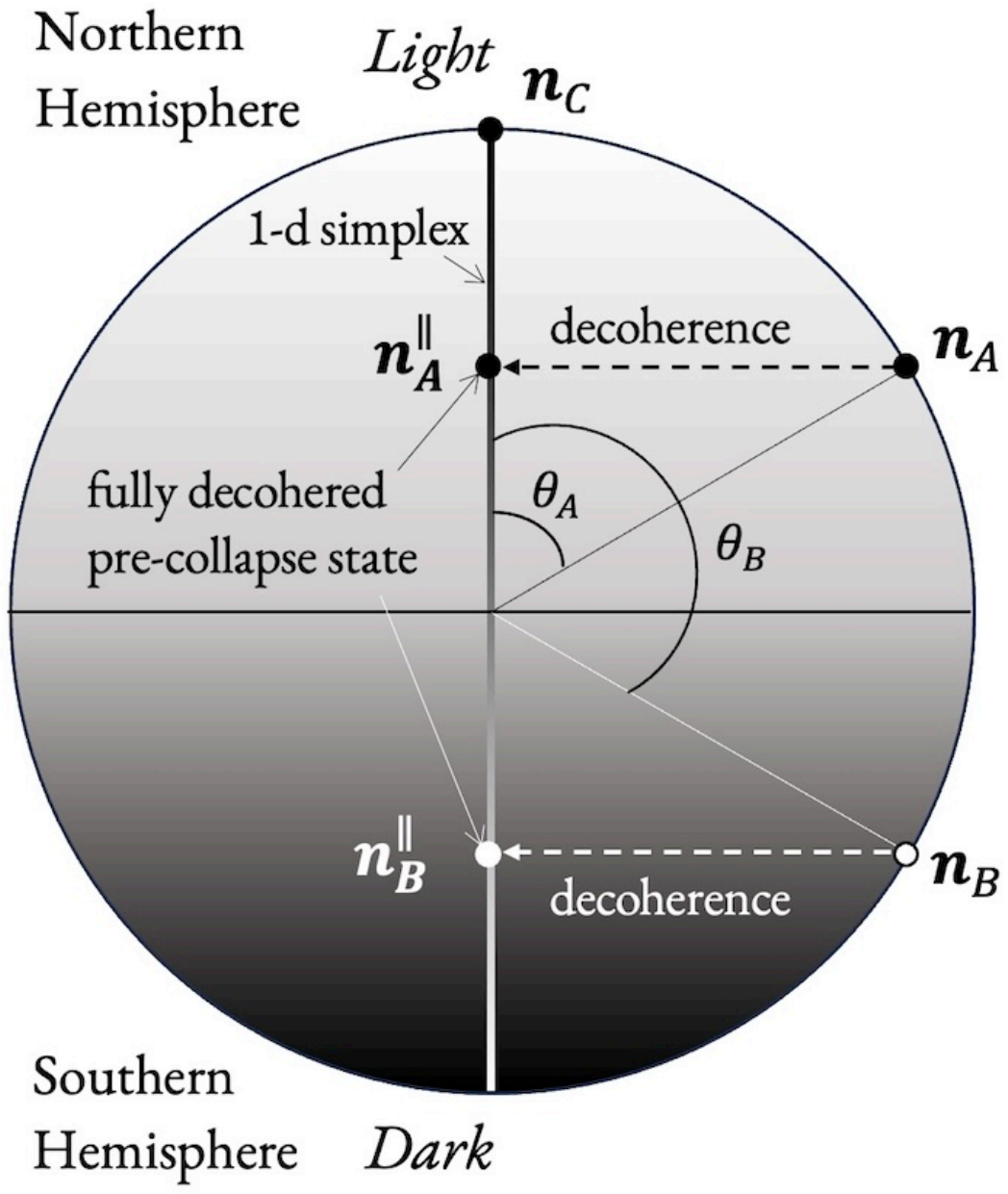

**Figure 11** A representation of the three pre-measurement vector-states (stimuli) $\boldsymbol{n}_A$, $\boldsymbol{n}_B$ and $\boldsymbol{n}_C$, at the surface of the Bloch sphere, and the associated decohered states (percepts) $\boldsymbol{n}_A^{\parallel}$ and $\boldsymbol{n}_B^{\parallel}$, inside the Bloch sphere, obtained by plunging them orthogonally with respect to the South-North axis (since $\boldsymbol{n}_C$ is an eigenstate of the measurement, it lies already on that axis, hence $\boldsymbol{n}_C = \boldsymbol{n}_C^{\parallel}$). Here we have assumed for simplicity that all vectors lie on a same plane of the sphere, and their polar angles are: $\theta_A = \frac{\pi}{3}$, $\theta_B = \frac{2\pi}{3}$ and $\theta_C = 0$.

So, we will use two different notions of distance, one that considers the circular arc between two vector-states, at the surface of the Bloch sphere, and the other one which considers the Euclidean distance between operator-states, inside the Bloch sphere. Hence, the normalized to 1 distance between the two decohered states $\boldsymbol{n}_A^{\parallel}$ and $\boldsymbol{n}_B^{\parallel}$ will be given by (see Figure 11):

$$d_{density}(\boldsymbol{n}_A^{\parallel}, \boldsymbol{n}_B^{\parallel}) = \frac{1}{2}\,|cos\,\theta_B - cos\,\theta_A|$$

Equipped with these two notions of distance, let us now explain why the phenomenon of categorical perception is naturally expressed in a quantum measurement process, when described in the extended Bloch representation. As we mentioned, the points forming the measurement simplex, in our case the sphere's diameter where the elastic band is stretched, is where the percepts lie, for the measurement in question. These percepts represent an "expected reality," which however is not to be understood as a collection of wild speculations, but rather as the best picture of what is considered to be real in relation to the meaning carried by the measured entity. This expected reality contains biases, and the warping of categorical perception is one of them.

More precisely, the perceiver is actively involved in the measurement process and her/his "expected reality" will play a crucial role in what happens, together with the reality of the measured entity, expressed by its pure state, which is instead measurement independent. At the level of the expected reality, the model is Kolmogorovian, the probabilities being an expression of the perceiver's lack of knowledge. This is realized in the extended Bloch model by the unpredictable point $\boldsymbol{\lambda}$ where the elastic will break; see Figure 10(d). Also, the connection between the reality of the considered entity (the vector-states describing the stimuli) and the elements of the expected reality (the on-simplex operator-states describing the percepts for the given measurement), is illustrated by the deterministic orthogonal "fall" of the point particle representative of the stimulus onto the elastic band (the 1-dimensional simplex), transforming it into a percept, as per Figure 10(b)-(c), thus reaching a stage where an outcome (an answer) is actualized, going back to a vector-state [see Figure 10(d)-(f)].

To see how a quantum measurement brings about the warping effect, let us consider, to fix ideas, a situation where only two colors exist, *Light* and *Dark*, so we are precisely in a situation that can be described in a three-dimensional Bloch sphere. Note that Eleanor Rosch formulated the rationale for the prototype theory for concepts while teaching colors to a primitive community in Papua New Guinea, whose language, called *Berinomo*, has just two names for colors (Rosch 1973). Let us locate the first color, *Light*, at the North Pole of the Bloch sphere, and the second color, *Dark*, at its South Pole. At the equator, the transition from *Light* to *Dark* will then occur (see Figure 11). Let us then introduce three different vector-states, the first one described by the unit vector $\boldsymbol{n}_A$, with polar angle $\theta_A = \frac{\pi}{3}$, the second one by the unit vector $\boldsymbol{n}_B$, with polar angle $\theta_B = \frac{2\pi}{3}$, and the third one by the unit vector $\boldsymbol{n}_C$, located exactly at the North Pole (hence, it is the eigenstate describing *Light*, with a polar angle $\theta_C = 0$), assuming for simplicity that they all lie on a same plane. When a *Light-Dark* color-measurement is performed, the pre-measurement states $\boldsymbol{n}_A$ and $\boldsymbol{n}_B$ deterministically transform into the *fully decohered pre-collapse operator states* $\boldsymbol{n}_A^{\parallel}$ and $\boldsymbol{n}_B^{\parallel}$, obtained by plunging the associated point particles into the sphere, orthogonally with respect to the one-dimensional simplex subtended by the *Light* and *Dark* outcome states, which is the region of the percepts, i.e., of the "expected reality" relative to this specific color-measurement.

The states $\boldsymbol{n}_A$ and $\boldsymbol{n}_B$ are, respectively, in the *Light* and *Dark* hemispheres of the Bloch sphere, hence they are two different colors, and we can now easily see that their transformation to the corresponding decohered states exhibits a warping that is a dilation. Indeed, their distance is $d_{pure}(\boldsymbol{n}_A, \boldsymbol{n}_B) = \frac{1}{3}$, i.e., one third of the maximal distance between two stimuli, while the associated percepts, corresponding to the decohered quantum states $\boldsymbol{n}_A^{\parallel}$ and $\boldsymbol{n}_B^{\parallel}$, have a distance $d_{density}(\boldsymbol{n}_A^{\parallel}, \boldsymbol{n}_B^{\parallel}) = \frac{1}{2}$, , i.e., one half of the maximal distance between two percepts. On the other hand, if we consider the stimuli $\boldsymbol{n}_C$ and $\boldsymbol{n}_A$, belonging to the same color, the opposite warping occurs. Indeed, on the stimuli side, the distance is again $d_{pure}(\boldsymbol{n}_A, \boldsymbol{n}_C) = \frac{1}{3}$, whereas on the percepts side we now have $d_{density}(\boldsymbol{n}_A, \boldsymbol{n}_C) = \frac{1}{4}$. This means that a warping takes place which is now a contraction. And this shows how the phenomenon of categorical perception is built into a quantum measurement.

If our hypothesis turns out to be correct, it will mean that we not only face a bias at the interface where human perception takes place, but likewise a similar bias would be present at the many interfaces where quantum measurements are executed in the physics' laboratories, and such a possibility opens up the following question: "What is the nature of the original pre-biased reality?"

## 4.3 Explaining quantum entanglement

In Section 2.3, we observed that *quantum entanglement* can be understood as a situation of *interconnected quantum entities*, where correlations can be created out of their connections, when joint measurements are performed. We emphasized that the mystery of quantum entanglement is not in a possible mechanism of *creation of correlations*, as it is easy to conceive macroscopic examples, like the historic vessels of water model, where by breaking the wholeness of a composite entity, correlations "not of the Bertlmann socks kind" can be created and used to violate Bell's inequalities. The real mystery of quantum entanglement is in the nature of the *connective element* that makes two otherwise spatially separated entities behave as if they were forming a single whole.

The conceptuality interpretation offers an extremely simple and persuasive answer to elucidate this mystery. The connective element at the origin of the quantum entanglement's connections is the very *substance of meaning*. Indeed, if it is true that quantum entities are conceptual entities, then it is also to be expected that they can share meaning, i.e., that they can *connect through meaning*, and that a cognitive entity interacting with them will be sensitive to the presence of these *meaning connections*, when confronted with an interrogative context. This also explains why quantum entanglement is the default state, in the sense that whenever entities are allowed to interact, they will typically entangle, so that entangled states are much more common than non-entangled ones (although that doesn't mean that they are always long lasting). Indeed, when two conceptual entities are brought together, in a same cognitive situation, if they do share meaning, then a meaning-connection will automatically be present, and the bipartite system they form will be described by means of an entangled state.

Take as an example a cognitive psychology experiment that was conducted and analyzed in our group (Aerts et al. 2018b,c). Participants were asked to select pairs of spatial directions that they considered to be the best example of *Two different wind directions*. In other words, participants were interrogated about two *Wind direction* conceptual entities, when in the state described by the conceptual combination *Two different wind directions*. This was done by giving the participants concrete examples of pairs of winds directions, from which they had to pick the pair they considered to be the best representative of the *Two different wind directions* state; see Figure 12.

To analyze the results, as is done in typical Bell-test experiments, four different joint measurements were proposed to the participants, using different combinations of the eight (cardinal and ordinal) directions on a *compass rose*. We skip here the details of how this was exactly done and refer the interested readers to Aerts et al. (2018b,c). What is important here to observe is that the statistics of the obtained results was able to violate Bell-CHSH inequalities with the same order of magnitude as in typical experiments with micro-entities in *singlet states*. And this means that human minds perceive wind directions to be connected through meaning in a similar way as measuring apparatuses, like Stern-Gerlach apparatuses, perceive quantum entities (like electrons or photons) to be entangled in the physics' laboratories, suggesting that meaning-like connections could also be in force for them. Note that the experiment we have mentioned, with the wind directions, is just an example of numerous experiments that have been carried out over the years, all showing that meaning-connections can generally violate Bell's inequalities and reveal the existence of genuine quantum-like correlations. See for example Aerts & Sozzo (2011) and Aerts et al. (2019), and the references cited therein.

At this point, one might ask: "Why in the laboratory we cannot detect the meaning-connection that would be at the origin of quantum entanglement, whereas we can detect the entangled entities?" In other words: "If photons (and other micro-entities) are meaning-entities, which can connect through meaning, why photons can be detected, but not the meaning substance forming their connections?" A possible answer, within the conceptuality interpretation, is that the meaning-element connecting two conceptual entities would be more abstract compared to the entities it connects, hence more *distant* (in a conceptual sense) from the spatiotemporal layer where the measuring apparatuses are located, and in that sense less accessible in being directly detected.

Consider again the example of the *Wind directions*. The connective element creating the correlations is here the concept *Different*. Clearly, the presence of it, implies that the directions of the two winds cannot be independent form one another, precisely because they are in a state such that they must be different. And within our human culture, *Different* is a much more abstract concept than the concept *Wind direction*, and in that sense less close to the conceptual entities that can leave traces in our spatiotemporal theater.

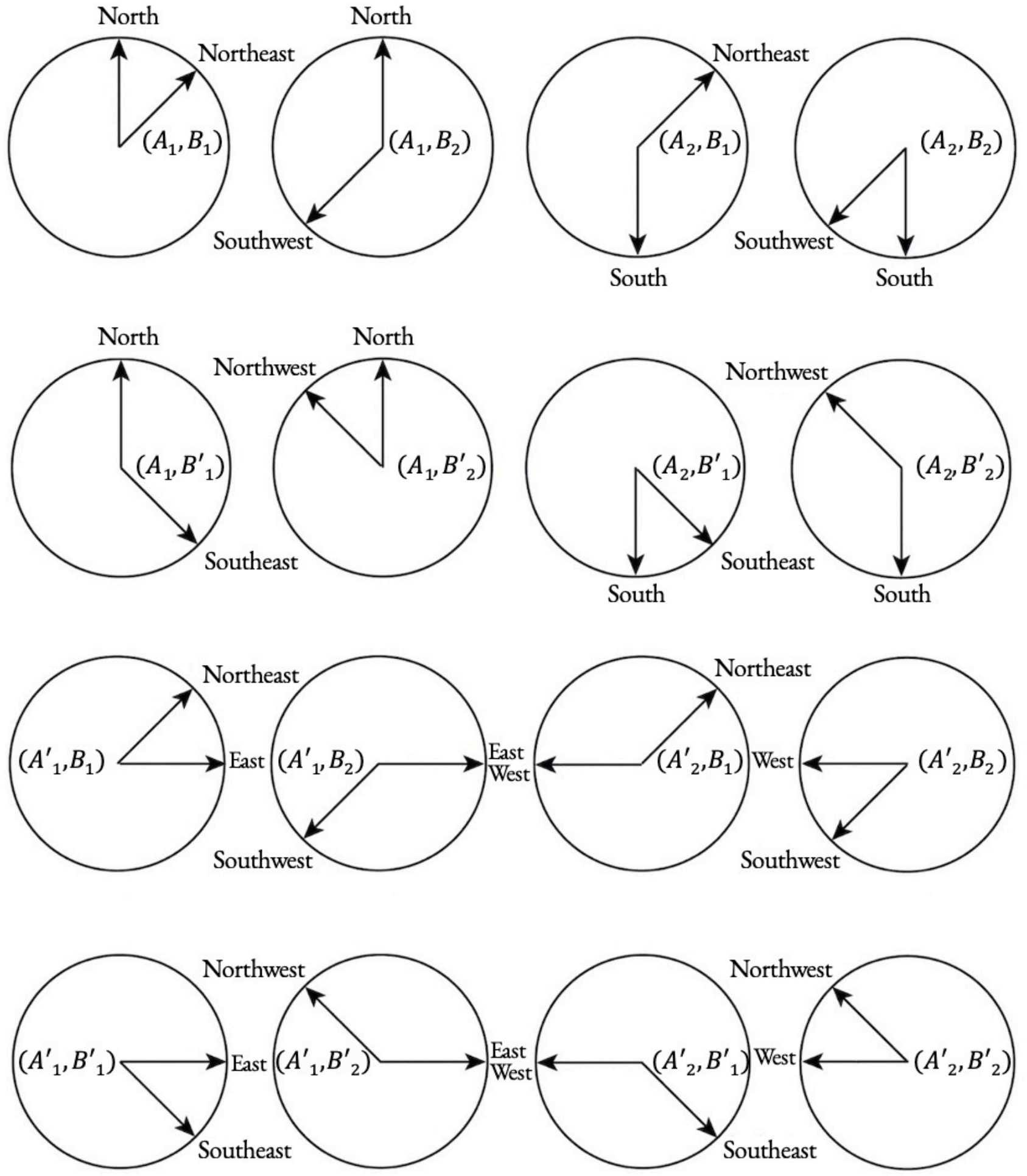

**Figure 12** The four pairs of winds directions that the participants in the experiment were allowed to select, for each of the four joint measurements considered, $AB$, $AB'$, $A'B$ and $A'B'$, which resulted in a violation of Bell's inequalities (Aerts et al. 2018b,c). More precisely, joint measurement $AB$ is the combination of measurement $A$ with measurement $B$, where $A$ consists in selecting either the North wind direction, or the South wind direction, and $B$ consists in selecting either the Northeast wind direction, or the Southwest wind direction. Hence, joint measurement $AB$ consists in asking a human participant to choose one among the following four couples of different wind directions: $(A_1, B_1)$ = (North, Northeast),

$(A_1, B_2)$ = (North, Southwest), $(A_2, B_1)$ = (South, Northeast) and $(A_2, B_2)$ = (South, Southwest), and similarly for the remaining three joint measurement: $AB'$, $A'B$ and $A'B'$.

A reflex of this greater abstractness of the entanglement connection can also be seen in the quantum formalism, when observing that the dimensionality of the *state of the connection* is greater than the dimensionality of the states it connects. To make full sense of the previous sentence, the extended Bloch representation needs to be applied to the description of entangled entities (Aerts & Sassoli de Bianchi 2016). Without going into details, let us assume for simplicity that the two entangled entities, $S_A$ and $S_B$, are two spin-$\frac{1}{2}$. When they are experimentally separated and are not considered to be part of a bipartite system, their states can be described by unit vectors $r_A$ and $r_B$, respectively, each belonging to its own three-dimensional Bloch sphere. But when they are considered as a composite entity, the Bloch sphere of states becomes 15-dimensional, and their overall state $r$ is a unit vector having a *tripartite direct sum structure*

$$r = \frac{1}{\sqrt{3}} r_A \oplus \frac{1}{\sqrt{3}} r_B \oplus r_{corr}$$

where $r_{corr}$ is a 9-dimensional vector describing the state of the connection between $S_A$ and $S_B$, whose components cannot be generally deduced from the components of the individual states $r_A$ and $r_B$ (unless the bipartite system is in a product state). So, we see that the quantum formalism, and more specifically its Blochean extension, also indicates that the entanglement connection happens on a dimensional level that is higher than that of the entities it unites, compatibly with the hypothesis that the meaning it would be associated with is more abstract (see also the discussion in the next section).

It should also be remarked that a singlet spin state is a very special state, in the sense that it contains no specific spin directions, as is clear that it is a rotationally invariant state. This means that it cannot be associated with well-defined individual spin properties prior to the measurements. This mirrors the situation of the conceptual entity in the *Two different wind directions* state. Indeed, here as well, the state describes an entity without specific wind directions, which cannot be associated with well-defined individual wind properties prior to the measurements, hence, like for the singlet state, it describes a more abstract situation.

One might wonder if there is a difference between the connection produced by entanglement and the connection produced by "ordinary" physical forces, like the *Coulomb force*, that can be described as a potential term in the Hamiltonian. Indeed, following the conceptuality interpretation, if meaning is what brings entities together, then one would expect ordinary forces, which also bring entities together, to themselves be associated with meaning connections. This is certainly compatible with the view of the conceptuality interpretation, if we think that, considering for instance the case of the *electromagnetic force*, one can describe its action as an exchange of *virtual photons*, i.e., of *virtual bosons*, and as we mentioned in Section 3, bosons are to be considered the natural building blocks of the protolanguage of the material world.

One could argue, however, that the ontological status of virtual bosons is not the same as that of real bosons, hence it would remain debatable if, say, a proton and an electron in a hydrogen atom are truly *communicating* when exchanging virtual photons. Without entering here into the debate of the virtual particles, we can just observe that when there is a force between two entities, then, following the conceptuality interpretation, one expects a situation of entanglement to emerge between the two entities interacting through the force in question. This because in the conceptual realm all connections are meaning-connections, and the latter are supposed to manifest as quantum entanglement. And indeed, that is exactly how things are. Take the case of the *hydrogen atom*. One is used to solve the problem of finding its stationary states (the eigenstates of the Hamiltonian) by introducing the *relative* and *center-of-mass* coordinates. When one does so, the center-of-mass and relative degrees of freedom appear to be separated, i.e., there is no entanglement between them. When instead one considers the coordinates of the electron and protons, the electron-proton bipartite system will in fact be entangled, as is clear that the evolution of the electron state, precisely because in interaction with the proton, cannot be independent from the state of the latter, hence the electron-proton system cannot be in a product state (Tommasini et al. 1998).

A last aspect worth mentioning in relation to entanglement is the following. As we said, human conceptual entities can easily violate Bell's inequalities. But they also typically violate the so-called *marginal laws*, which in physics are usually referred to as *no-signaling conditions*. The reason of this designation is that when they are obeyed, then a superluminal communication exploiting the entanglement phenomenon is impossible, although their violation does not necessarily imply that such superluminal communication can be achieved (this is an aspect that is often not emphasized enough). For example, the above psychological experiment, with the *Two different wind directions*, does violate the marginal laws, although not in a very pronounced way. Now, it is generally believed that in the standard quantum formalism the marginal laws are always satisfied. This is true whenever the joint measurements used in the formulation of Bell's inequalities are represented by product self-adjoint operators, relative to the tensor product Hilbert space modeling the states of the composite entity. Hence, one might consider the violation of the marginal laws as a reason for rejecting the notion of *meaning-connection* as a candidate to explain quantum entanglement.

In that respect, two things should be said. The first one is that there is no a priori reason for the joint measurements to be represented by product self-adjoint operators. It is considered appropriate to do so because we look at the polarizers or Stern-Gerlach used to test Bell's inequalities as entities that, in their macroscopicity, are experimentally separated when they are at a sufficient spatial distance from each other, in a typical Bell test experiment. But is this really the case? Perhaps it is precisely the interaction with a quantum entity prepared in a singlet state that reveals that the apparatuses in question are not really separated, from an experimental point of view, despite their spatial separation. In this regard, it is interesting to note something that is not very well-known: that the marginal laws are constantly violated in Bell-test experiments (Adenier & Khrennikov 2007, De Raedt, Michielsen & Jin, 2012, De Raedt & Michielsen 2013, Adenier & Khrennikov 2017, Kupczynski 2017).

Considering that the standard quantum formalism, with its a priori prescription of describing joint measurements as product measurements (Aerts et al. 2019), requires the marginal laws to be always obeyed, the tendency is to believe that a violation of the latter in Bell-test experiments would just be due to errors. On the other hand, the conceptuality interpretation predicts the possibility of these violations, hence, they may not turn out to be experimental errors, but the result of cognitive-like mechanisms involved in the very creation of the correlations. In other words, they may turn out to be a confirmation of the validity of the conceptuality interpretation. Note that a description of entanglement in a Bell-type situation where joint measurements are also considered to be entangled was elaborated in Aerts & Sozzo (2014).

The second thing to say is that one should not expect the human conceptual realm to exhibit the same level of sophistication as the proto-conceptual realm of the material entities. Indeed, the latter being much more ancient, it is reasonable to expect that it evolved its language up to the point that it became much more regular, with more recurring patterns, etc. (think for example of the language used by the operators in a control tower, when communicating to airplane pilots during their landing and take-off). And this, among other things, would also explain why waves and interference phenomena are so effective in describing cognitive overextension and underextension effects, which in the human domain typically appear when people judge conjunctive or disjunctive concepts (Hampton 1988, Aerts et al. 2020).

So, the fact that in human cognition the marginal laws are violated, whereas they usually are supposed to not be violated in the physical world, cannot be considered as a valid counterargument against the conceptuality interpretation.

## 4.4 Explaining quantum uncertainty

The next mystery on our list is quantum uncertainty. As we observed in Section 2.4, it does not lie in the fact that measurements can be mutually incompatible, so that they cannot be jointly executed, and if sequentially executed their order will affect the final statistics of outcomes. In the human cognitive domain, interrogative processes are also, generally speaking, mutually incompatible, as the phenomenon of *question order effects* clearly highlights (Busemeyer & Bruza 2012, Aerts & Sassoli de Bianchi 2017a).

Quantum uncertainty, or quantum complementarity, becomes a real conundrum for those who adhere to a purely spatiotemporal vision of our physical reality, when they observe that also observables like position and momentum are complementary and that, different from the wooden cube example that we described in Section 2.4, we cannot have both properties sharply actual at the same time. This means that quantum uncertainty touches a deeper aspect of the reality of a quantum entity.

If an electron would jointly have a well-defined position and momentum, at a given moment, one would be able to apply to its state the classical laws of motion and determine its trajectory. But we know quantum entities cannot be associated with spatiotemporal trajectories. So, the question is: "Why micro-entities, like photons, electrons, atoms, etc., cannot be in states where both position and momentum are jointly maximally sharp or, to put it another way, they cannot typically be in states where position and momentum are jointly maximally *un*sharp?" (Remember that reverse uncertainty relations also exist, although state-dependent).

The answer, within the conceptuality interpretation, is rather straightforward: if it is true that the micro-entities are conceptual in nature, they cannot be in states that are maximally abstract and at the same time maximally concrete. And since a maximally concrete state is considered to describe a situation of maximal localization in space, whereas a maximally abstract state is considered to describe a situation of maximal de-spatialization, Heisenberg's uncertainty principle appears to be built in the conceptuality interpretation. And the same goes for the reversed version of Heisenberg's relations (Mondal et al. 2017), as is clear that if the micro-entities are conceptual in nature, they cannot either be in states that are minimally abstract and at the same time minimally concrete.

One might object that the reversed relations do not have the same status as the standard relations, since the upper limit is not fixed but depends on the state. Therefore, it is in principle possible to construct states that make this limit arbitrarily large.‡ One can respond to this objection in several ways. First, by observing that localization in physical space only partially captures the classical notion of an object, with which we usually associate the idea of a state of maximum concreteness. In fact, in classical physics, an object possesses not only a well-defined position but also a well-defined momentum. Therefore, we cannot expect the *position-momentum* binomial to be a faithful translation of the more fundamental *concrete-abstract* binomial. Another observation is that the interpretation of quantum uncertainty as resulting from a tradeoff between concreteness and abstractness might only apply for that important class of concepts that are structured around typical or idealized examples, called *prototypes*, which would then be described by single gaussian-like wave functions, for which upper limits for the uncertainty relations would also exist. Certainly, the analysis of how the concrete-abstract binomial is reflected in the quantum formalism still requires careful investigation, also taking into account the fact that the Hilbertian formalism is not necessarily able to capture the full complexity of a pancognitivist reality.

Speaking of *abstract versus concrete* states of a concept, this is the right moment to bring the reader's attention on a possible confusion, which we mentioned already in Section 4.2, due to the fact that, on one hand, we have the human cognitive activity, and on the other hand, we have the physical world *per se*, which according to the conceptuality interpretation also participates in cognition, but at a different level in our reality. This means that there can be differences between conceptual domains in the way the abstract versus concrete direction should be interpreted.

We humans have abstracted most of our concepts from what we imagine to be the objects of our everyday experience. For instance, by interacting with real spatiotemporal dogs, at some point we have extrapolated from these experiences the *idea of dog*, to describe not a single specific animal, but what the essence of these animals are, the properties they share, independently from their individual differences. So, at some point, a concept was created, the *Dog* concept, which is a *one-word concept*. Such one-word concept is like the *title of a story*; the story of all the interactions humans have had with the different dogs they met. A story that explains what dogs do, how they move, what they eat, how they look like, how they interact with each other, with other animals, with humans, etc. It would be very impractical (very inefficient), when referring to that collection of experiences, to tell their whole story every time; but since these are shared experiences, it is sufficient to use a single-word-concept (or a few-words-concept) to evoke them all. This is like when, to evoke the story contained in a book, we simply mention its title.

The above distinction is important because in the conceptuality interpretation one considers that there is a fundamental line for going from the most abstract concepts to the most concrete concepts (see Figure 13), and such line is characterized by the fact that, quoting from Aerts et al. (2020):

> [...] the more abstract concepts are those that are expressible by single words, and concreteness increases when the number of conceptual combinations increases, so that the most concrete concepts are those typically described by large aggregates of meaning-connected (entangled) single-word concepts, which is what in our human realm we would generically indicate as stories, like those written in books, articles, webpages, etc. We don't mean here stories only in the reductive sense of novels, but in the more general sense of clusters of concepts that are combined together in an interesting way, so as to create a well-defined meaning.

‡ For example, if we consider the superposition of four Gaussian wave packets $|\psi\rangle = \frac{1}{\sqrt{4}}\left(|\psi_{q,p}\rangle + |\psi_{q,-p}\rangle + |\psi_{-q,p}\rangle + |\psi_{-q,-p}\rangle\right)$, where $|\psi_{q,p}\rangle$ is centered on $q$ in position space and on $p$ in momentum space, and similarly for the other three components, then it is not difficult to show that $\Delta Q \Delta P \sim qp$, as $q, p \to \infty$. The reason for this is that the above state is formed by the superposition of four components with a fixed spread, i.e., with a fixed *standard deviation* ($\sigma$ in position space and $1/2\sigma$ in momentum space), which are placed at arbitrarily large distances. Thus, the total spread, both in momentum and position space, can become arbitrarily large.

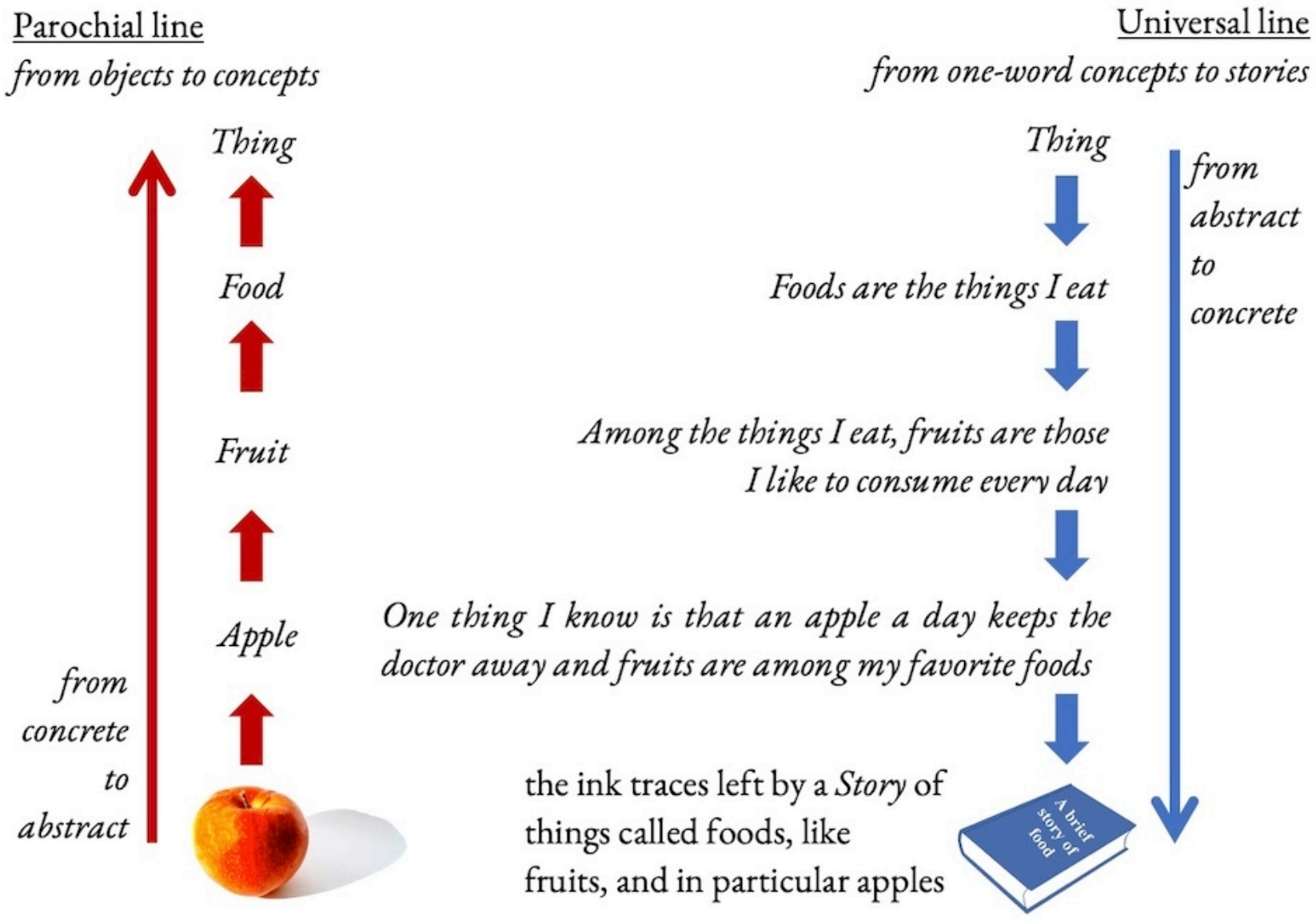

**Figure 13** (color online). In the human conceptual domain, there are two main lines connecting *abstract* to *concrete*. The first one goes from *concrete* to *abstract* (left): from *objects* to *collections of objects having common features*. The second one, more fundamental, goes from *abstract* to *concrete* (right): from *one-word concepts* to *stories*, formed by the combination of multiples concepts.

Of course, we must distinguish the title of a book from its meaning content. The title refers to the story in the book, but being formed by a few words, it is a much more abstract concept than the story it refers to, which is a much more concrete conceptual entity within the human conceptual domain (proof of that, it is not unusual to find books with the same title, while there are no cases of authors who have written the exact same book). Now, these two conceptual domains, the human and the physical, they come into contact when a person tells the story of what a dog is, and as s/he does so, s/he looks in the direction of a specific dog (see Figure 14). That dog specimen is also a story, but not the story the human is telling. It is the story of all the conceptual atomic and molecular entities forming its body. And this is a story that is not made by combining human concepts, but by combining those concepts that are the quantum entities; a story that a human mind is obviously not able to understand.

One might ask at this point if in the conceptual-cognitive domain of the physical world there would be an equivalent of the title of a story. A possible answer is that this would correspond to the emergent properties of a large aggregate of conceptual entities. If we consider a macroscopic entity with sufficient internal cohesion, allowing it to manifest as a unitary body (instead of a collection of unconnected fragments), it can interact with similar macroscopic entities according to dynamics that primarily interest their wholeness and not the details of what happens inside of them, at the level of their constituents. We can think of a ball interacting with pins. The dynamics occurs at the level of the macroscopic bodies, in the sense that similar dynamics would be produced even if, for example, the material of the ball in question were to be modified (a wooden ball is able to land pins similarly to a plastic ball, if they have similar masses). In other words, also in the physical domain there would be different cognitive layers, whose dynamics unfold based on different meaning-interactions. In the human cognitive domain, we can think of how a bookseller positions the newcomer books in the shelves of her/his bookstore. Obviously, s/he does not have to read the new entries to classify them: s/he can simply look at their *emerging information*, like the titles and the authors, and based on it, s/he will be able to place the different books in one shelf instead of another, regardless of the details of the stories they tell.

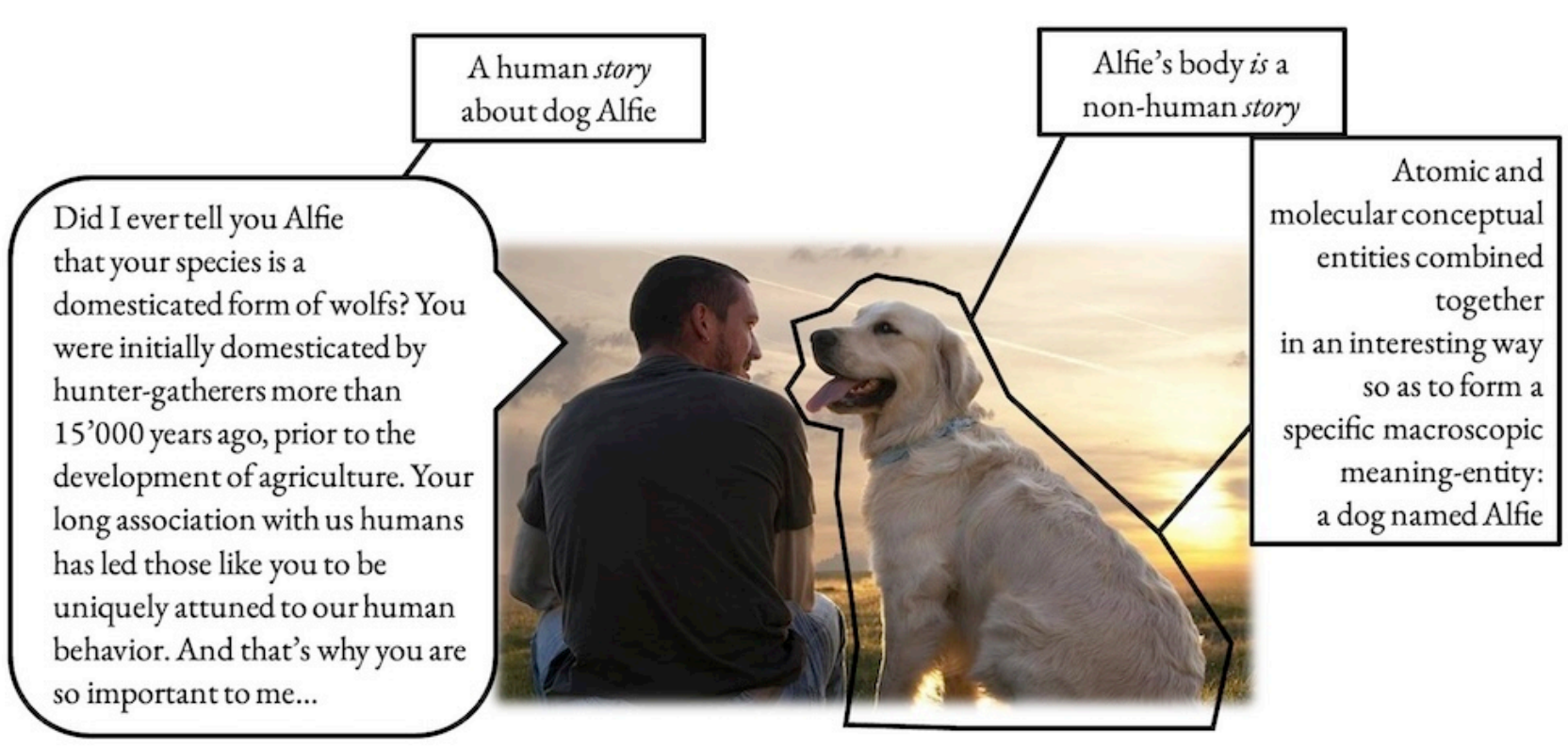

**Figure 14** (color online). A dog named Alfie can be associated with two stories. One is the story his owner can tell, about dogs in general and Alfie in particular; the other is the story "written" in the aggregate of atoms and molecules that make up Alfie's body, as a macroscopic physical entity. These two stories belong to two completely different cognitive domains, even though they refer to the same entity.

Speaking of books, it also worth observing that we can go in a bookstore and buy *book A and book B*. This is so because *book A* is an object, *book B* is an object, and *book A and book B* is also an object: a composite one. But we cannot go into a bookstore and buy *book A or book B*. There is no object that can be associated with the term "book *A* or book *B*". In a quantum mechanical language, *book A or book B* is a macroscopic entity in a superposition state, and we do not usually find it in our everyday physical reality. This doesn't mean, however, that it would be an impossible entity. As we mentioned already in Section 2.2, in controlled laboratory conditions we have already been able to put relatively large physical entities, like organic molecules, in superposition states (Gerlich et al. 2011, 2013).

Does the conceptuality view allow us to understand why we do not find macroscopic entities in a superposition state? To answer this question, let us reason again within the human conceptual domain. To avoid any confusion, it is good to point out that a book is not a "human story" conceptual entity; a book only contains the *traces* left by it (see the discussion in Section 4.1). As proof of this, a same story can be published on different paper volumes, of different sizes, or in electronic format, etc.

Now, as we emphasized already when we talked about the dog, a book is where two stories of a very different kind meet: the one written by the human author of the book, whose traces are left in the book-material-entity, and the story that the book *is*, being an entity formed by fermionic and bosonic quantum entities, which according to the conceptuality interpretation are essentially conceptual in nature. Therefore, reasoning within the human conceptual realm, the question one has to ask is: "Why don't we find the traces of the entity *story A or story B* in bookstores (or anywhere else), contrary to the case of the entity *story A and story B*?

To put the above question in different terms, we know that the conceptual realm is closed with respect to the conjunction (and) and the disjunction (or) connectives: if *A* and *B* are concepts, then *A and B* and *A or B* are both also concepts. On the other hand, the realm of objects is only closed with respect to the conjunction connective. Indeed, the conjunction of two objects is still an object: a composite object formed by the juxtaposition of the two objects in question; but the disjunction of two objects is not an object anymore. So, can the interpretation that equates objects with stories help us clarify this difference between concepts and objects?

Here we must understand that what we call "object" is a limit case of a concept, hence an object would still be a conceptual entity. In the same way, also a human story is a limit case of a concept: it is a complex combination of one-word-concepts combined according to all sorts of meaning bonds, expressed at different levels, i.e., at different scales in the story. Take the story that Erwin Schrödinger once told, about what life is, whose traces can be found in his famous booklet entitled *What is life?* (Schrödinger 1944); and take the story that Alan Alexander Milne once told, about a teddy bear, whose traces can be found in his equally famous booklet entitled *Winnie-the-Pooh* (Milne 1926). If we can all be certain that nobody ever told the story *What is life? or Winnie-the-Pooh*, it is because the ambiguity introduced by the connective *or*, between these two stories, would be meaningless in our human culture, and we humans only tell stories to convey meaning, not to convey an absence of meaning. Consequently, no publisher ever printed a book such that the first pages contain (the ink traces of) Schrödinger's story, then there is a subsequent page with the printed word "or" and finally the remaining pages are those with (the ink traces of) Milne's story. You will never find such a book in a bookstore, although it is certainly not impossible to create it.

Note that the fact that we have written and published this article makes the existence of such strange book a little more probable. A Zen Master of the Rinzai school could for instance decide to create it after having read our writing, as a tool to lead some of his students to gain *kensho* (awakening), by trying to capture its meaning, as is traditionally done with the study of *kōan*.

The comparison of the two conceptual domains which are the human and the physical, allows us to gain further insight on the articulation between the abstract and the concrete, which in turn allows us to better understand the difference between non-spatial and spatial physical states. About this, we studied in Aerts (2013) the nature and frequency of occurrence of *conjunctions* of concepts compared to the nature and frequency of occurrence of *disjunctions* of concepts, on the World-Wide Web. In this way, we found a systematics that gives us a foretaste of how our notion of *physical space filled with objects* might have emerged from a conceptual framework.

If we combine couples of concepts that we have chosen more or less at random, such as *Car* and *Building*, *Flute* and *Bass*, *Horse* and *House*, *Table* and *Sun*, Yahoo searches on the World-Wide Web for their combinations, using the connective element "and", i.e., *Car and building*, *Flute and bass*, *Horse and house*, *Table and sun*, show that these are more common than the combinations using the connective element "or", i.e., *Car or building*, *Flute or bass*, *Horse or house*, *Table or sun*. As we have seen, the connective element "or" introduces an abstraction, and considering our understanding of quantum uncertainty as due to the interplay between abstraction and concreteness, this means that when the connective "or" is introduced between two concepts, a superposition state is formed, which is less localized than the states forming the superposition. In contrast, the connective element "and" generally introduces more concreteness. This means that when used between two concepts, a state is formed that is more localized. And since our experiments with the World-Wide Web show that, for arbitrarily chosen concepts, the longer the combination the more common the "and" connective element becomes, compared to the "or" connective element, this indicates the general tendency of the former to localize texts on the Web pages.

This localization process necessarily stops at the level of the final cognitive product, i.e., the concretely accessible pages of the World-Wide Web. So, if we take the latter as an example of a cognitive environment, these final (concrete) Web pages are for human cognition the equivalent of what *ordinary matter that fills space* is for the physical reality. Within a classical view of the latter, matter is assumed to fill space, and more generally spacetime, by giving rise to objects. This should be considered as the limit of a process of objectification which, however, never truly allows physical entities to reach the status of *objects as such*. In this sense, the notion of object is only an idealized one, which plays an essential role solely in the idealized theory that is classical physics. This is confirmed by the fact that ordinary matter is never truly localized, as it contains atoms and molecules and within these substructures particles are in non-localized superposition states.

We encounter a similar situation in the domain of human cognition. Indeed, the "or" connective element, which gives rise to non-localized states, appears frequently in the form of little *meaning molecules* on the pages of the World-Wide Web. The examples we have identified are of the following type: *The window or the door*, *Laugh or cry*, *Dead or alive*, *Coffee or tea*, etc. These are the equivalents, for the human cognitive domain, of what molecules and atoms of ordinary matter are for physical reality.

A notion that allows us to understand more easily the above, is that of *bifurcation*. Indeed, if one focuses on the evolution in time, a superposition state introduces a situation of *contextual bifurcation*, with the collapsed states describing the different bifurcation branches (and more generally, *multifurcation* branches), the number of which depends on the dimension of the Hilbert space (Aerts & Durt 1994, Aerts & Sassoli de Bianchi 2014, 2016). The *meaning molecule* given by the disjunction *Coffee or tea*, for example, can be seen as the situation of a table where people are ready to consume something and are offered coffee or tea, with the bifurcation corresponding to their possible choices. We can easily understand that a person at that table who has not yet chosen coffee or tea, holds both possibilities into existence, and that *this* corresponds to a genuine *element of reality*.

When we look at the microworld, we know that the superposition states are very abundant, which means that our physical reality also holds into existence such elements of bifurcation. Therefore, there is no "cloud of electrons" around the nucleus of an atom, but only electrons that have not yet made themselves available to participate in a measurement process capable of actualizing a place in space for them. That this situation is much more common in the microworld than in the macroworld is a well-known observation, associated with one of the unsolved quantum problems:

*Why is the macroworld mostly classical, while the microworld is generally quantum?*

According to the conceptuality interpretation, if this is still considered to be a problem it is because we make a subtle mistake: we compare the behavior of pieces of fermionic matter, on the surface of the planet, with the behavior of elementary entities such as electrons. These macro-pieces of fermionic matter, like stones, tables, chairs and human bodies, behave very similarly to how we expect objects to behave, being mostly localized in space and separated from each other. But as our "coffee or tea" disjunction example suggests, we shouldn't compare the electrons interacting with the nucleus of an atom with these macro-pieces of fermionic matter, if we describe the latter as mere inert objects, but we could do so if we embed them in a dynamic that includes bifurcations, as is clear that even a stone that lies inert for a long time along a path can suddenly be picked up by a walker and be given a new place in her or his personal collection of stones.

This is already much truer for tables, and certainly for chairs, which regularly change places, thus experiencing constant bifurcations, not to mention the body of a living being, continually subjected to choices that guide its state along possible alternatives. In humans, we place the origin of these choices in their brains (and more generally in their minds), and when in the conceptuality interpretation we claim that cognition also takes place in the microworld, we do not mean it in the sense of the existence of specific cerebral structures and associated mental faculties. The body-mind split and associated brain structures are probably specific to how cognition organized itself and emerged in the animal and human kingdoms. In fact, the way in which cognition is more specifically organized in the microworld is the subject of further investigation in the conceptuality interpretation, whose main ingredients are, on the one hand, the existence of concepts and the linguistic structures that emerged from their combinations and, on the other, the existence of cognitive apparatuses capable of understanding their meaning and making choices accordingly (see also the discussion of Section 5).

Let us further elaborate on the macroscopic-classical versus microscopic-quantum issue, because the conceptuality interpretation contains at least part of an explanation for this situation. Consider a gas, such as that formed by the air molecules of our planet's atmosphere. These molecules, when at room temperature, constantly collide with each other at the speed of a jet plane, on average. A calculation can then show that their *de Broglie wavelength* is far too small to cause their wave functions to overlap, to allow quantum interference, and that is why air at room temperature behaves like a classical gas. If we lower the temperature in a lab, bringing it very close to absolute zero, a new state of the gas can be realized that is called a *Bose-Einstein condensate*, at least if the molecules are bosons and if we can keep the gas dilute enough (Cornell & Wieman 2002, Ketterle 2002).

What happens is that when the temperature is significantly reduced, the average velocity slows down so much that the de Broglie wavelengths become so large that the different molecules' wave functions can overlap and give rise to quantum interference effects, and other quantum phenomena (Aerts 2014, Aerts & Beltran 2020). In other words, a gas that at room temperature is dominated by randomness becomes a coherent structure at extremely low temperatures. (Note that the temperatures that physicists can achieve in their labs are much colder than the coldest places that have ever existed in the universe and so, in this sense, this is probably a new reality situation for our cosmos).

What is the way to interpret the above within the conceptuality interpretation? We usually think of air molecules at room temperature as entities that are always "close" to each other, but is this really a correct perspective? They collide with each other constantly at very high average speed, but their collisions are totally random, which is also why we consider air, as a whole, as inert matter. What would be an equivalent situation when considering people communicating with each other? That of a place where people shout random words at each other without even listening to what others are saying, and in that situation, we are certainly not tempted to say that those people would be "close" (in a communicative sense) to each other. On the contrary, we would say that there is lack of communication. This means that we should consider the air of our planet as an ensemble of molecules that are so distant from each other that they effectively behave as separated entities, so much so that quantum effects become negligible, the reason for this being the presence of too much heat and the consequent destruction of coherence that these same entities could create in colder situations.

But is it even true that classical mechanics makes it completely impossible to describe a typical quantum uncertainty situation? In a previous section, we already gave the example of a pencil standing on its point to illustrate the classical notion of symmetry breaking and bifurcation. And indeed, in classical mechanics, it follows from the mathematical formalism that in addition to states of *stable equilibrium* there are also states of *unstable equilibrium*. The latter contain a type of quantum uncertainty, although limited by the mathematical formalism of classical mechanics, which does not allow such states to become *contextual*, i.e., capable of reacting in different ways with respect to different measurements. On the other hand, the Hilbert space formalism of quantum mechanics allows to make such states of unstable equilibrium genuinely contextual, describing them as superposition states.

However, it is important to note that the Hilbert space formalism also carries a fundamental structural limitation, it does not allow the description of (experimentally) *separated physical entities.* The formalism capable of describing both contextually unstable states and separated systems was developed using mathematical notions (*lattices* and *closure structures*) which for the time being are still little known by current quantum physicists, so that further developments of this more general formalism have not yet taken place (Aerts 1982b).

A related issue is that of the *linearity* of the Hilbert space formalism. Indeed, it would be curious, to say the least, that such a specific mathematical formalism as that of a *complex linear vector space* could describe the physical world in all generality. Considering what we have just mentioned, namely that the simplest situation of two separated entities cannot be described in a Hilbert space, we suspect that the success of the Hilbertian framework should only be considered in relative terms. In this regard, it is worth mentioning that, for example, it is possible to show that the Born rule emerges as a *universal average* over probabilities that are not necessarily Hilbertian (Aerts & Sassoli de Bianchi 2015c, 2017b).

It would undoubtedly be useful, and enlightening, thinking about future developments of the conceptuality interpretation, if a mathematical formalism could be more fully developed that could remedy the structural shortcomings of the standard Hilbertian approach (Aerts 1982b), where linearity could, for example, be introduced as an additional possibility only when the physical situation really requires it (Aerts & Sassoli de Bianchi 2015c, 2017b).

### 4.5 Explaining quantum indistinguishability

The last of the quantum features we have to explain is *quantum indistinguishability*. As we mentioned in Section 2.5, its true mystery lies in the fact that, contrary to Leibniz's ontological principle of the identity of the indiscernibles, two quantum entities can be genuinely indistinguishable and nevertheless still be (in some sense) individuals. If this appears as an impossibility for objectual entities, it is instead entirely expected for conceptual entities. Indeed, when a message does not convey any element of distinction between two conceptual entities, then for the receiver of that message they are genuinely indistinguishable.

We must remember here that when we use our language to describe a given number of conceptual entities, the content of our description is a specification of the state of these entities (Aerts, Sassoli de Bianchi & Sozzo 2016). If nothing in the description can be used to differentiate them, then, by definition, they will be in a *state of indistinguishability* and, cognitively speaking, they will behave accordingly.

For example, if we say *Eleven animals*, all the *Animal* concepts in the multipartite *Eleven animals* state are in the exact same state, hence they all carry the exact same meaning. And the presence of the numeral-concept *Eleven* guarantees that we are not in the presence of a single entity, but of a collection of entities. To take another example, when in our bank account we have, say, *Eleven thousand euro*, we are again in a situation of genuine indistinguishability, as proved by the fact that when we ask the bank clerk *One thousand euro*, we have no way of telling her/him which of them s/he has to withdraw from the account. Things would obviously be different with banknotes, which for example contain distinct serial numbers, allowing them to be distinguished. Here again we have an example of the important difference between a conceptual entity, like the entity *Money* is, and the traces that it can leave in our spatiotemporal theater, for example in the form of banknotes.

Now, if the indistinguishability of quantum entities is a consequence of their conceptual nature, plus the fact that they can be in states conveying no information that cognitive entities can use to distinguish them, and considering that in the physical domain indistinguishability has profound consequences on the statistical behavior of collections of identical entities, a prediction of the conceptuality interpretation is that one should also find non-classical (non-Maxwell-Boltzmann) statistical behaviors within the human conceptual domain (Aerts, Sozzo & Veloz 2015). And indeed, this prediction finds confirmation in the analysis of the structure of human language.

To show this, we follow Aerts & Beltran (2020) and consider a collection of one-word-concepts that are meant to convey meaning through a specific narrative, i.e., a story. We can call these one-word-concepts forming the story *cognitons*, in analogy with *photons*, the quanta of the electromagnetic field. The cognitons of the story are in states that are specified by the different words that are used to tell it. Some words are repeated very frequently, other less frequently, and repeated words are to be understood as cognitons in a same meaning-state. More precisely, let us assume that the story in question is formed by $N$ cognitons and that $N_0$ of them are in the state corresponding to a given word, say the word *The*, which is the most often repeated; $N_1$ of them are in the state corresponding to another word, say the word *Be*, which is the second word the most often repeated in the story in question; and so on, with $N_n$ cognitons being in the state corresponding to, say, the word *Anthropomorphization*, which is the one less often repeated. To these $n$ different words one can associate different energy levels, $E_i = i, i = 0, \ldots, n$, then ask what a good function would be for modeling the observed occupation numbers $N_i \equiv N(E_i)$.

According to the conceptuality interpretation, bosons are the natural building blocks of languages, hence, a story told in human language should behave similarly to a *gas of confined bosons*, the bosons in question being here the cognitons, i.e., the one-word-concepts that can occupy the different energy levels, corresponding to the different words used in the story. To put it another way, the conceptuality interpretation predicts that a *natural language story* should behave similarly to a boson gas of entangled words, at a given temperature, and it is perhaps one of the most notable confirmations of its validity to observe that, indeed, the Bose-Einstein distribution

$$N_{BE}(E_i) = \frac{1}{A\, e^{\frac{E_i}{B}} - 1}$$

with the parameters $A$ and $B$ determined by the conditions

$$N = \sum_{i=0}^{n} N_i \qquad\qquad E = \sum_{i=0}^{n} N_i\, E_i$$

offers a remarkable fit of the data (Aerts & Beltran 2020). On the other hand, the Maxwell-Boltzmann distribution

$$N_{MB}(E_i) = \frac{1}{C\, e^{\frac{E_i}{D}}}$$

with the parameters $C$ and $D$ to be again determined by the above conditions on the total number $N$ of words and their total energy $E$, offers a very poor fit of the data; see Figure 15.

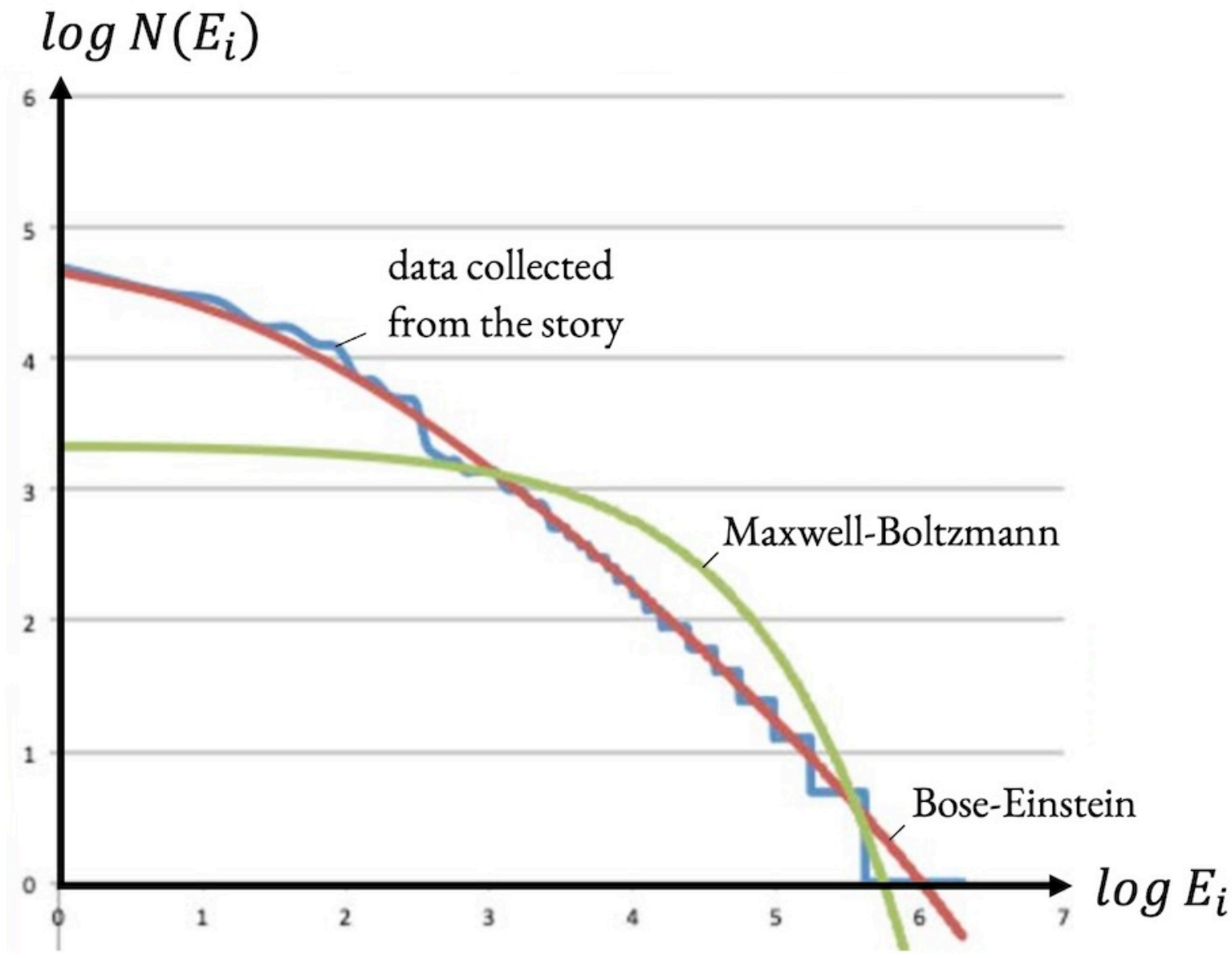

**Figure 15** (color online). The (*log* of) the number of words' appearances $N_i$ in the *Winnie the Pooh* story entitled *In Which Piglet Meets a Heffalump* (Milne 1926), as a function of (the *log* of) the associated energy levels $E_i$. The blue graph represents the data directly collected from the story to be modeled. The green graph corresponds to the Maxwell-Boltzmann distribution, $N_{MB}(E_i)$, and the red graph to the Bose–Einstein distribution, $N_{BE}(E_i)$. The red and blue graphs coincide almost completely, while the green graph does not coincide at all with the blue graph of the data. The Figure is adapted from Figure 1(b) in Aerts & Beltran (2020).

But that is not all. If one allows the modeling to become more realistic, by letting the spacing between the consecutive energy levels not to be a constant, as it is the case for the simple situation of the quantum harmonic oscillator, but to vary according to a more general power law

$$E_i = i^p \Delta E + E_0$$

as it happens for more general shapes of the confinement potential (Bagnato et al. 1987), then an almost perfect match of the data can be obtained (Aerts & Beltran 2020). These results are even more significant if one considers that they offer what is probably the first convincing explanation of the origin of *Zipf's law* (Zipf 1935, 1949), i.e., the fact that in a *corpus of natural language* the most frequent word will occur approximately twice as often as the second most frequent word, three times as often as the third most frequent word, and so on. In ultimate analysis, this unexpected behavior would result from the fact that human concepts, similarly to bosonic entities, behave as indistinguishable entities, as evidenced by the fact that the words that are repeated in a story can be exchanged without affecting its meaning.

This is no longer the case if, in a printed book containing the ink traces of such story, we physically exchange some of the words printed on paper, for example cutting and pasting them in the new places. The reader of the book will certainly notice the manipulation that has taken place, though it will not affect the meaning of the story that the book tells, since only words that are the same, i.e., that have the same meaning, are exchanged. And this is the reason why printed words obey the Maxwell-Boltzmann statistics, whereas the non-printed words, the cognitons, obey the Bose–Einstein statistics, from which Zipf's law follows.

# 5 Conclusion

Paul Bush (2001) once asked:

> Is quantum theory 'nothing more' than a statistical theory, or could the referent of a quantum state be a single object? This question reflects a division of the quantum physics community into two distinct 'cultures', corresponding to the two options it addresses. The task of justifying an interpretation that goes beyond a minimal probabilistic interpretation amounts to the task of solving all the quantum puzzles and paradoxes that have made the foundations of quantum mechanics such an exciting enterprise throughout the history of this theory. The needs of quantum cosmology, and also the most recent developments in the experimentation with single microsystems, certainly encourage the search for a sound individual interpretation.

In this article, we defended the thesis that not only the quantum puzzles and paradoxes can be addresses, but that they can also be satisfactorily solved. But to do so, the prejudice that quantum mechanics, and physics in general, would be about objects, must abandoned. Quoting also from de Ronde & Massri (2016):

> [...] it is not obvious nor self-evident that we must presuppose this specific 'object-property' metaphysical scheme to interpret the quantum formalism. Unfortunately, the main discussions in the literature [...] presuppose implicitly the notion of object. This metaphysical choice has produced many interpretational problems. Obviously, such problems cannot escape their own presuppositions, and exactly because of this reason, QM has been confined to a discussion within the limits of this very specific metaphysical perspective.

In accordance with de Ronde & Massri, we have proposed a non-objectual ontology and metaphysics for quantum mechanics, in accordance with the observation that quantum entities are non-spatial, and more generally non-spatiotemporal, and that spatiality is usually regarded as a key feature of physical objects. By closely observing the behavior of the quantum entities, then using the experience we have accumulated over the years about the effectiveness of the quantum modeling of cognitive processes (*quantum cognition*), it became possible not only to say with relative certainty what a quantum entity *is not* (it is not a particle, it is not a wave, it is not a field, it is not a spatial entity, it is not an object) but also to propose what a quantum entity *is*, i.e., what its *nature* is. Our proposal is that a quantum entity has a conceptual nature, i.e., a nature which is equivalent to that of a *human concept*, as understood in the ambit of *human cognition*, and we hope to have succeeded in providing the reader with sufficient elements in favor of our thesis, and sufficient stimuli to encourage the further investigation of its explanatory power.

We are of course aware that other quantum puzzles need to be addressed and explained, in addition to those we have evoked in this work. If we have not done so here it is only to avoid overstretching an already long text. In other words, we had to make choices, and we refer the interested reader to Aerts (2009, 2010a,b, 2013, 2014), Aerts & Sassoli de Bianchi (2018), Aerts et al. (2020), Aerts & Beltran (2020, 2022) and Aerts & Aerts Arguëlles (2022) for a discussion of other experimental situations, like the *double-slit experiment*, the *delayed-choice experiment*, the *quantum eraser experiment* and the *phenomenon of quantization* (which we briefly addressed in Sec. 4.2). We also refer to Aerts et al. (2020), and Aerts & Sassoli de Bianchi (2023), to appreciate the fecundity of the conceptuality interpretation in shedding light on the not well understood aspects of *relativity theory*. Furthermore, the *pancognitivist* framework that emerges from this interpretation has also strong implications on our understanding of *evolution* and the appearance of complex life forms on the surface of our planet, and we refer to Aerts & Sassoli de Bianchi (2018) for more details on that.

Within this concluding discussion, it is important to consider a possible objection: namely, that macroscopic objects, made of ordinary matter, could not be understood as cognitive entities, since cognitive activity typically requires a brain and sensory organs to be carried out. We briefly addressed this issue in Aerts et al. (2020), recalling that "if it is true that a certain behavior presupposes a certain organization, this doesn't mean that a same organization would be needed to obtain a same behavior." The above objection is also known among those who study *plant sentience* and defend its possibility as a proper non-metaphorical attribute. So, mutatis mutandis, part of the literature addressing the issue of *cognition in plants* is also relevant for the issue of *cognition in inert matter*, but it is beyond the scope of this work to deal in a satisfactory way with this multifaceted issue.

Let us here just emphasize that one should not assume that a cognitive entity is also, necessarily, a conscious entity. Also, a general theory of consciousness and cognition should avoid as much as possible any anthropocentric bias, like the one saying that *no brain* implies *no cognition*. In that respect, Hiernaux (2021) proposes to generally understand consciousness as a type of cognitive activity, and cognition as a type of behavior. So, without the need of having brain-like structures, an entity would be able to be cognitive, in a minimal sense, if it can behave in a non-automatic and non-predictable way, i.e., in a way that involves the presence of decision processes over different alternatives. And such minimal view on cognition is certainly compatible with the central hypotheses of the conceptuality interpretation.

On the question of the meaning of the conceptuality interpretation in relation to the nature of our physical universe, and apart from the subject of consciousness, there are of course several possibilities. One of them is that, simply, cognitive interactions, and the corresponding cognitive structures, would turn out to be the most organizationally efficient, and this would explain why the universe would have begun to structure itself in such a way. Another is that *Mind* would be the fundamental nature of the universe, with matter and energy mere manifestations of it. Whatever the meaning of the universe, if the conceptuality interpretation proves to be true, the universe would then be like a big "talking shop," where chatter is constantly going on everywhere.

In concluding, it is instructive to briefly clarify how the conceptuality interpretation relates to other interpretations of quantum mechanics. A comprehensive analysis of such comparisons would require a dedicated article and here we shall only succinctly evaluate the conceptuality interpretation in light of Maudlin's (1995) three fundamental claims, commonly employed to distinguish and categorize interpretations. The first claim pertains to the completeness of the wave function, asserting that it fully determines the state of a system, by specifying all its actual and potential properties. From the viewpoint of the conceptuality interpretation, this assertion is only partially accurate. Although a Hilbert space vector often adequately captures the totality of a system's properties, there are notable exceptions. Specifically, genuine states necessitate a description also through density operators, for example during measurements or whenever systems are entangled (Aerts & Sassoli de Bianchi 2016). Moreover, the interpretation does not inherently privilege the Hilbert space geometry, as empirical evidence from human cognition reveals violations of relationships implied by this geometry and by Born's rule, notably the *marginal laws* (Aerts et al. 2019). Hence, the linear structure of Hilbert space is understood merely as an approximation of a more encompassing formalism capable of also describing phenomena such as separation (Aerts et al. 2024a).

Maudlin's second claim concerns the exclusively deterministic evolution of the wave function, typically described by the unitary evolution dictated by Schrödinger's equation. This determinism is not upheld by the conceptuality interpretation. Within a pancognitivist framework, cognitive and decision-making processes pervade reality, implying that evolution predominantly manifests as a nonlinear process, open to a diversity of outcomes at any given moment. Unitary evolution thus emerges merely as a particular limiting case within a broader spectrum of evolutionary possibilities, corresponding to scenarios where there is only one possible outcome.

Finally, addressing Maudlin's third claim, which emphasizes genuinely indeterministic measurement contexts characterized by multiple possible outcomes – only one of which materializes according to the relative frequencies prescribed by the Born rule – the conceptuality interpretation essentially agrees with its validity. Indeed, it affirms the existence of a single actualized outcome at each measurement. However, as previously noted, Born's rule itself is to be understood only as an approximation within a more general probabilistic framework (Aerts & Sassoli de Bianchi 2014, 2015b,c).